\documentclass[printer]{aa}

\usepackage{graphicx, xcolor}
\definecolor{cerulean}{rgb}{0.0, 0.48, 0.65}
\usepackage[colorlinks, citecolor=cerulean]{hyperref}
\usepackage{pdflscape}

\newcommand{\kms}{km\,${\rm s}^{-1}$}
\newcommand{\Msun}{M$_{\odot}$}
\newcommand{\Lsun}{L$_{\odot}$}
\newcommand{\Rsun}{R$_{\odot}$}
\newcommand{\Teff}{$T_{\rm eff}$}
\newcommand{\logg}{$\log\,g$}
\newcommand{\vsini}{{$\varv\,\sin i$}}
\newcommand{\vmac}{$\varv_\mathrm{macro}$}
\newcommand{\vmic}{$\varv_\mathrm{micro}$}
\newcommand{\halpha}{H$\alpha$}
\newcommand{\hbeta}{H$\beta$}
\newcommand{\ve}{$v_{\rm{e}}$}

\defcitealias{Bodensteiner2020a}{Paper~I}
\defcitealias{Bodensteiner2021}{Paper~II}
\usepackage{graphicx}

\usepackage{txfonts}

\begin{document}

   \title{The young massive SMC cluster NGC~330 seen by MUSE
          \thanks{Full Table A.1 is only available in electronic form
at the CDS via anonymous ftp to cdsarc.cds.unistra.fr (130.79.128.5)
or via https://cdsarc.cds.unistra.fr/cgi-bin/qcat?J/A+A/.}
          \thanks{Based on observations collected at the ESO Paranal
          observatory under ESO program 60.A-9183(A) and 0102.D-0559(A).}}
    \subtitle{III. Stellar parameters and rotational velocities}

   \author{J. Bodensteiner\inst{1,2},
   H.~Sana\inst{2},
   P.~L.~Dufton\inst{3},
   C.~Wang\inst{4},
   N.~Langer\inst{5},
   G.~Banyard\inst{2},
   L.~Mahy\inst{6},
   A.~de Koter\inst{2,7},
   S.~E.~de~Mink\inst{4,7},
   C.~J.~Evans\inst{8},
   Y.~Götberg\inst{9},
   V. H{\'e}nault-Brunet\inst{10},
   L.~R.~Patrick\inst{11},
   F.~R.~N.~Schneider\inst{12}
    }
   \institute{European Southern Observatory,  Karl-Schwarzschild-Strasse 2, Garching by Munich, Germany\\
    \email{julia.bodensteiner@eso.org}
     \and
     Institute of Astronomy, KU Leuven, Celestijnenlaan 200D, 3001 Leuven, Belgium
     \and
     Astrophysics Research Centre, School of Mathematics and Physics, Queen's University Belfast, Belfast BT7 INN, UK
     \and
     Max-Planck-Institut für Astrophysik, Karl-Schwarzschild-Straße 1, 85741 Garching, Germany
     \and
     Argelander-Institut für Astronomie, Universität Bonn, Auf dem Hügel 71, 53121 Bonn, Germany
     \and
     Royal Observatory of Belgium, Avenue Circulaire 3, B-1180 Brussels, Belgium
     \and
     Astronomical Institute Anton Pannekoek, Amsterdam University, Science Park 904, 1098 XH, Amsterdam, The Netherlands
     \and
     European Space Agency (ESA), ESA Office, Space Telescope Science Institute, 3700 San Martin Drive, Baltimore, MD 21218, USA
     \and
     The observatories of the Carnegie institution for science, 813 Santa Barbara Street, Pasadena, CA 91101, USA
     \and
     Department of Astronomy and Physics, Saint Mary’s University, 923 Robie Street, Halifax, NS B3H 3C3, Canada
     \and
     Departamento de Astrof\'{\i}sica, Centro de Astrobiolog\'{\i}a, (CSIC-INTA), Ctra. Torrej\'on a Ajalvir, km 4, 28850 Torrej\'on de Ardoz, Madrid, Spain
     \and
     Heidelberger Institut für Theoretische Studien, Schloss-Wolfsbrunnenweg 35, 69118 Heidelberg, Germany
    }
   \date{Received xx Month Year; accepted xx Month Year}

  \abstract
   {The origin of the initial rotation rates of stars, and how a star's surface rotational velocity changes during the evolution, either by internal angular momentum transport or due to interactions with a binary companion, remain open questions in stellar astrophysics.}
   {Here, we aim to derive the physical parameters and study the distribution of (projected) rotational velocities of B-type stars in the $\sim$35\,Myr-old, massive cluster NGC~330 in the Small Magellanic Cloud. NGC~330 is in an age range where the number of post-interaction binaries is predicted to be high near the cluster turnoff (TO).}
   {We develop a simultaneous photometric and spectroscopic grid-fitting method adjusting atmosphere models on multi-band Hubble Space Telescope (HST) photometry and Multi Unit Spectroscopic Explorer (MUSE) spectroscopy. This allows us to homogeneously constrain the physical parameters of over 250 B and Be stars (i.e., B-type stars with emission lines), brighter than $m_\mathrm{F814W}=18.8$~mag.}
   {
   The rotational velocities of Be stars in NGC~330 are significantly higher than the ones of B-type stars. The rotational velocities vary as a function of the star's position in the color-magnitude diagram, qualitatively following predictions of binary population synthesis. A comparison to younger clusters shows that stars in NGC~330 rotate more rapidly on average.}
   {
The rotational velocities of the $\sim$35\,Myr old population in NGC 330 quantitatively agree with predictions for a stellar population that underwent significant binary interactions: the bulk of the B-type stars could be single stars or primaries in pre-interaction binaries. The rapidly spinning Be stars could be mass and angular momentum gainers in previous interactions, while those Be stars close to the TO may be spun-up single stars. The slowly rotating, apparently single stars above the TO could be merger products. The different \vsini-characteristics of NGC~330 compared to younger populations can be understood in this framework.
   }


    \titlerunning{The young massive SMC cluster NGC 330 seen by MUSE. III.}
   \authorrunning{Bodensteiner et al.}

   \maketitle
%

\section{Introduction}\label{Sec:intro5}
Apart from the initial mass, the initial rotation rate is considered to be one of the key properties with major influence on the evolution and end-points of (massive) stars \citep[e.g.,][]{Maeder2000a, Brott2011, Ekstrom2012, Woosley2006}. Rotation can drive internal mixing processes that can have several effects on the star: new fuel is mixed into the core, the main-sequence (MS) lifetime is increased, the luminosity is significantly altered, and nuclear processed material might be brought to the surface \citep[see e.g.,][]{Maeder2000a, Howarth2001}. Several mixing mechanisms are thought to be at play in massive stars, including semi-convection, convection, thermohaline mixing, and mixing induced by gravity waves \citep[e.g.,][]{Rogers2013} or rotation \citep[for example by shear or meridional circulations, see e.g.,][]{Maeder2000a}. One-dimensional evolutionary models show that the more rapidly a star is rotating at birth, the stronger the impact of rotation-driven internal mixing \citep{Brott2011}. The effectiveness and importance of rotational mixing have, however, been questioned and alternative mechanisms to change the surface spin and surface abundances of stars have been proposed \citep[e.g.,][]{Hunter2007, Almeida2015, Aerts2019}. 

High surface rotation can result both from single- and binary-star evolution:  through angular momentum redistribution from the core to the envelope during the MS \citep[see e.g.,][if the initial rotation is sufficiently high]{Ekstrom2008, Hastings2020}, or through accretion of angular momentum during mass transfer \citep[e.g.,][]{deMink2013} or by tides \citep{deMink2009a}. 
Recent binary population synthesis predictions by \citet{WangC2020} showed that, combined with the position in the Hertzsprung-Russell diagram (HRD), the rotation rate of stars is one of the main tracers of a possible binary history \citep[see also][]{WangC2022}. This is most obvious for mass gainers through case-B mass transfer \citep{Kippenhahn1967}, which are expected to be spun up to high rotation rates. THhose might observationally appear as classical Be stars \citep[e.g.,][]{Pols1991, Hastings2021}, which are rapidly rotating B-type stars surrounded by a circumstellar decretion disk that leads to strong emission lines in their spectra \cite[e.g.,][]{Rivinius2013}. In contrast, merger products are thought to be slowly rotating as they are expected to spin down quickly after an initial puffed-up stage \citep[][and references therein]{Schneider2020}. 

Constraining the rotation rates of large, homogeneous samples of massive stars as a function of their metallicity, evolutionary stage and physical parameters greatly helps to identify the mechanism and evolutionary scenarios that are responsible for observed distributions of surface rotation velocities. Milky Way studies indicate that cluster stars rotate, on average, faster than field stars \citep{Slettebak1949, Abt2002, Strom2005, Huang2006, Daflon2007, Garmany2015}. These differences are mainly explained in two different ways: on the one hand, they are attributed to the star formation process, where stars in denser environments are thought to be formed with higher initial rotation rates than stars in less dense environments \citep[e.g.,][]{Strom2005, Wolff2007}. On the other hand, they are explained as an evolutionary effect where field stars are on average more evolved, and hence rotate slower as their radius has increased towards the end of the MS \citep[e.g.,][]{Huang2006, Huang2008}. Focusing on the fastest rotators, \citet{Huang2010} found that they are likely either formed by evolutionary spin-up of the most massive stars, or by binary mass transfer. \citet{Braganca2012} reported that the rotational velocities of runaway B-type stars in the field resemble those of cluster B-type stars, suggesting that they were born in clusters and later ejected from these denser environments.

Comparing rotational velocity distributions of massive OB stars in the Milky Way to those of the Large Magellanic Cloud (LMC) association N\,206 \citep{Ramachandran2018a} and the Small Magellanic Cloud (SMC) Wing, \citet{Ramachandran2019} showed that stars in the SMC have, on average, higher rotational velocities. This was attributed to two effects: firstly, lower-metallicity stars are more compact and, at a given mass, thus have generally higher rotational velocities \citep[e.g.,][]{Ekstrom2008}. Secondly, lower-metallicity stars have weaker winds \citep[if they exhibit winds in the first place,][]{Mokiem2007} such that they lose less angular momentum.

In the LMC, the VLT FLAMES Tarantula Survey \citep[VFTS,][]{Evans2011} homogeneously measured the rotational velocity distributions of OB stars in the young 30~Doradus region, a massive star forming complex. \citet{Ramirez-Agudelo2013} studied the rotation rates of about 200 apparently single O-type stars in 30 Dor. 
They found two components in the projected rotational velocity (\vsini) distribution. A majority of stars rotate at around 80\,\kms\, while there is a high-velocity tail of fast rotators. This tail is qualitatively similar to expectations from binary interaction products \citep{deMink2013}. Focusing on the primaries of O-type stars in binary systems in the same region, \citet{Ramirez-Agudelo2015} found that primaries in short-period systems are faster rotators than single O-type stars likely as a result of tidal synchronisation \citep[see also][]{Mahy2020a}. Furthermore they found a clear lack of very rapidly rotating binary O-type stars, in line with expectations that prior interaction is needed to produce such rapid rotation.

\citet{Dufton2013} showed that the 30 Doradus B-star \vsini-distribution is bi-modal, with one peak below 100\,\kms\, and a high-velocity component above 250\,\kms. Given that this study was performed in a cluster environment and no spatial or radial differences were found in the rotation rates, it seems unlikely that the bi-modality is due to observations of cluster and field stars, or stars formed in different star formation episodes. \citet{Hunter2008a} found the rotation rates for LMC B-type stars to be similar to the ones reported for Galactic B-type stars. 

Focusing on the $\sim$2-Myr old cluster NGC~346 in the SMC, \citet{Dufton2019} investigated the binary properties and rotational velocities of almost 250 O- and early B-type stars. While the authors found indications that the stars in NGC~346 are on average more rapidly rotating than Galactic early-type stars, the differences in the rotational velocity distributions are not statistically significant. The possible higher than average rotational velocity would be in line with the larger number of rapidly rotating Oe and Be stars detected in SMC clusters \citep{Martayan2007b, Martayan2007a}.

Photometric studies of slightly older clusters, that is with ages between $\sim$30\,Myr and 1\,Gyr reported on the presence of a split MS in the cluster color-magnitude diagrams (CMDs). One of the proposed explanations attributed this to different rotational velocities of stars in the two sequences \citep[e.g.][]{Bastian2009}, which was confirmed by spectroscopic observations \citep[see e.g.,][and Sect.\,\ref{subsec:vsini_others}]{Marino2018, Kamann2021, Kamann2023}. The origin of the different rotational velocities of the populations was, among other theories, explained as an imprint of previous binary interactions in those cluster \citep[see e.g.,][]{WangC2020, WangC2022, WangC2021, Sun2021}.

In this context, the intermediate-age, massive SMC cluster NGC~330 offers the possibility to obtain the rotational velocity distribution of stars in a low metallicity environment. Previously, NGC~330 was estimated to be 20\,Myr old \citep{Keller1999}, but more recent works found ages of around 35-40\,Myr \citep{Bodensteiner2020a, Eldridge2020, Patrick2020}. This puts it in an age range where recent theoretical computations predict a large number of binary interaction products to exist.

This is the third paper in a series. In \citet[][hereafter \citetalias{Bodensteiner2020a}]{Bodensteiner2020a} we described our multi-epoch observations on NGC~330 obtained with the Multi Unit Spectroscopic Explorer (MUSE) and characterized the massive star content. In \citet[][hereafter \citetalias{Bodensteiner2021}]{Bodensteiner2021} we measured radial velocities (RVs) and investigated the current binary properties of the massive star population. Here, we use MUSE spectra and HST photometry to quantify their physical properties. In Sect.\,\ref{sec:tlusty_fit} we introduce a newly implemented grid-based method to derive stellar parameters. We describe our resulting stellar parameters, in particular the rotational velocities, in Sect.\,\ref{sec:fitting_results}. In Sect.\,\ref{sec:otherworks} we compare our findings to previous works of clusters in both the LMC and SMC, and give a summary in Sect.\,\ref{sec:conclusions}.

\section{Determination of atmospheric and physical parameters}\label{sec:tlusty_fit}

Here, we describe the estimation of stellar parameters, namely the effective temperature \Teff, surface gravity \logg, projected rotational velocity \vsini\, (where $\varv$ is the stellar rotational velocity and $i$ is the inclination angle), radius $R$, and bolometric luminosity $L$. As a majority of the stars in the sample are hotter than 15\,000\,K, non-LTE effects may be important in their atmospheres \citep{Kudritzki1979, Kilian1994, Nieva2007}. We develop a grid-based algorithm based on the non-LTE \textsc{tlusty} BSTAR2006 grid \citep{Hubeny1995, Lanz2007}. The grid provides synthetic spectra for \Teff\, between 15000 to 30000\,K with steps of 1000\,K, and \logg\, from 3.0 to 4.75\,dex with steps of 0.25\,dex, where $g$ is expressed in cgs units. We use the grid computed for SMC metallicity (Z = 0.2\,Z$_{\odot}$), assuming a fixed microturbulent velocity of 2\kms. The synthetic spectra cover the UV to optical wavelength range (900--10000\,\AA).

Our newly developed fitting routine\footnote{available on github: \url{https://github.com/jbodenst/}} provides two modules to derive stellar parameters: \textit{1)} the optical co-added MUSE spectra are compared with rotationally broadened \textsc{tlusty} model spectra degraded to the appropriate spectral resolution and wavelength binning of MUSE (see Sect.\,\ref{subsec:specfit}), and \textit{2)} the available high-quality HST photometry from \citet{Milone2018} is compared to \textsc{tlusty} model SEDs to provide additional constraints on the radius of the star (see Sect.\,\ref{subsec:photfit}), which allows an estimate of the bolometric luminosity.

\subsection{Combination of MUSE spectra}\label{Sec:ObsAndSpec}

\begin{figure*}\centering
    \includegraphics[width=0.99\textwidth]{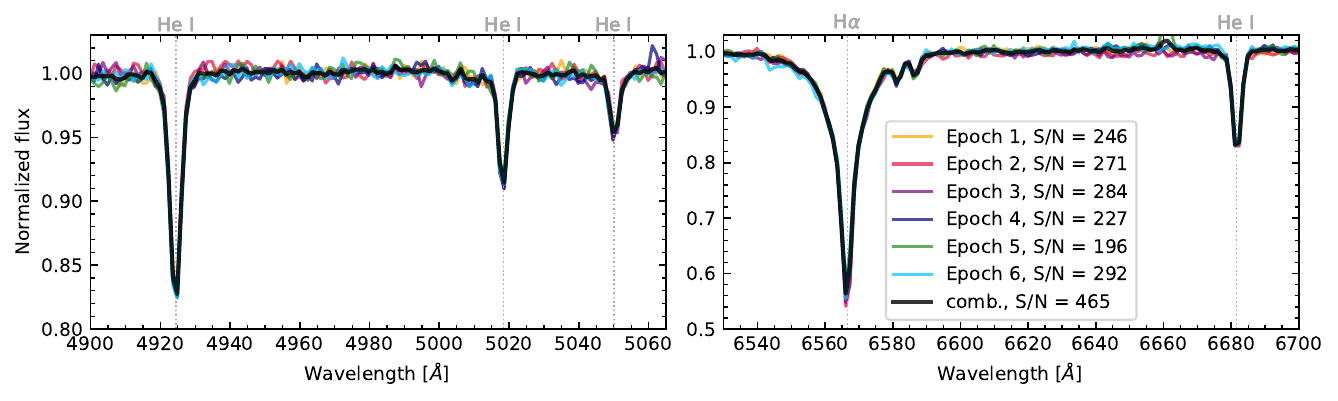}
    \caption{\label{Fig:combine_specs} Example of the co-addition of spectra for the apparently single star \#654. Individual epochs are shown in color (as indicated in the legend) while the combined spectrum is shown in black. The legend also indicates the measured S/N ratio at 6100\,\AA. Main spectral lines are marked.}
\end{figure*}
The MUSE observations, the reduction of the data including sky subtraction, the extraction of spectra and their normalization is explained in detail in \citetalias{Bodensteiner2020a}. We extracted four to six spectra (covering the wavelength range between 4600 and 9300 $\AA$ at a resolving power between 1700 and 3700) for 324 B and Be stars, six blue supergiants (BSGs) and ten red supergiants (RSGs) and measured RVs at each observed epoch for 282 of the B and Be stars and all of the BSGs (\citetalias{Bodensteiner2021}).

Knowledge about the RV variability of the stars allows us to shift the spectra to rest wavelength and stack them, weighted by the signal-to-noise (S/N) at 6100\,\AA{}. In the following, we focus on the 191 B and 83 Be stars. We omit eight stars with composite, variable spectra indicative of double-lined spectroscopic binary systems (SB2, found in \citetalias{Bodensteiner2021}). Among the 274 analysed stars, 26 are previously detected single-lined spectroscopic binaries (SB1s), where only one binary component can be identified in the spectra. Combining the spectra significantly increases their S/N ratio with typical values between 200 for the fainter and over 400 for the brighter B and Be stars. An example of the individual spectra of star \#654 (with individual S/N ratios between approximately 125 and 290 at around 6100 $\AA$) and the combined spectrum with a S/N of over 450 is shown in Fig.\,\ref{Fig:combine_specs}.

\subsection{Spectroscopic fit}\label{subsec:specfit}
The observed MUSE spectra allow us to constrain \Teff, \logg\, and \vsini\, of a star, which we do based on a grid search. For each model in the two-dimensional \textsc{tlusty} grid (\Teff\,-\,\logg), we apply the instrumental broadening corresponding to the MUSE (wavelength-dependent) resolving power, rebin the models to the MUSE wavelength step of 1.25\,\AA, and add rotational broadening using the python package \textsc{pyastronomy} \citep{pyasl}. We here consider rotational velocities ranging from 0 to 500\,\kms\, with steps of 20\,\kms, which is chosen to be lower than the typical accuracy of the \vsini-estimates as discussed below. We fit \Teff, \logg, and \vsini\, simultaneously as the former two cannot be fitted without taking rotational broadening into account, and vice versa. Additionally, the uncertainties can be correlated.

We then compare the observed spectrum to every model in the grid by computing the spectroscopic $\chi^2_\mathrm{spec}$ for each model as
\begin{equation}
    \chi^2_\mathrm{spec} =  \frac{1}{\mathrm{S/N}} \cdot \sum^{n}_{i=1} (\mathrm{obs}_i - \mathrm{model}_i)^2 \quad,
\end{equation}
where obs$_i$ and model$_i$ are the observed spectrum and the model at a given wavelength with index $i$, respectively, $n$ is the total number of spectral pixels considered, and S/N is the measured S/N ratio. We use the measured S/N rather than individual errors on each pixel of the spectrum as these errors are typically overestimated by the spectral extraction via the fitting of the point-spread function (PSF), especially for fainter stars.

We define six diagnostic regions in the MUSE spectral range (also shown in Fig.\,\ref{Fig:specfit_id654}): 
\begin{itemize}
    \item the blue part of the spectrum between 4600 and 5060 \AA, which contains several diagnostic lines such as \hbeta{}, four \ion{He}{i} lines at $\lambda\lambda$ 4713.15, 4921.93, 5015.68, 5047.74 \AA, and some metal lines of \ion{Si}{ii}, \ion{N}{ii}, \ion{C}{ii} and \ion{O}{ii} (which are in most cases too narrow and weak to be detected in MUSE),
    \item the region around the \ion{He}{ii} line at 5411 \AA,
    \item the region around \halpha{}, between 6500 and 6640 \AA,
    \item the \ion{He}{i} line at 6678.15 \AA,
    \item the \ion{He}{i} line at 7065.19 \AA, and
    \item the Paschen region between 8500 and 8930 \AA.
\end{itemize}

Often-used diagnostic lines for (early) B-type stars, such as the ratios between \ion{He}{i} and \ion{Mg}{ii}, or \ion{Si}{ii} and \ion{Si}{iii}, cannot be used as the lines are either too weak or outside the wavelength region covered by MUSE. The \ion{He}{i} line at $\lambda$ 5875.62\,\AA is also not available due to a \ion{Na}{i}-filter required when using MUSE with the laser-supported adaptive optics system \citepalias[see][]{Bodensteiner2020a}.

The regions containing the \halpha{} and Paschen lines were particularly affected by nebular emission, telluric absorption, and sky subtractions. Additionally the \halpha{} line in the observed spectra turns out to be systematically stronger than in the models, which might be caused by an over-subtraction of the sky. The Paschen region further suffers from identifying the true continuum. We therefore refrain from including these regions in the $\chi^2_\mathrm{spec}$-computation but use them as a consistency check when visually inspecting the fits. After visually inspecting every fit, we further exclude other regions for some stars when they appear noisy, contaminated, or badly normalized. The $\chi^2_\mathrm{spec}$ was computed in all remaining regions.

The estimation of rotational velocities based on the MUSE observations is mainly hampered by the relatively low resolving power, which varies between $R=1700-3700$ from the blue to the red part of the spectrum. This corresponds to a formal velocity resolution of about 150 to 80\,\kms, respectively. Previous studies based on MUSE data have, however, shown that a determination of \vsini\, with a $1\sigma$ accuracy of 30\,\kms\, is possible for cooler stars that provide a multitude of spectral lines \citep[e.g., ][]{Kamann2018b}. Here we show that it is also possible for hotter stars when using several spectra lines for our MUSE targets that have significant rotational broadening.

Given the availability of the \textsc{tlusty} models we have to adopt a microturbulent velocity \vmic=\,2\,\kms. As typical values of \vmic{} are of the order of a few \kms\, for B-type MS stars, this is much lower than the typical \vsini\, and thus a valid assumption. Furthermore we assume that the macroturbulent velocity \vmac{} is negligible. The macroturbulent velocity for B-type stars is below 100 \kms\, in most cases \citep[see e.g.,][]{Simon-Diaz2017}. While this is significant, it is lower than the typical rotational velocities and will be below or similar to the resolution capabilities of MUSE for most stars. The derived rotational velocities are thus upper limits and might be, on average, slightly overestimated. Our fitting approach also does not take into account the deformation of very rapidly rotating stars, which can have a strong impact on the derived stellar parameters \citep{Abdul-Masih2023}. The assumptions are further discussed in Sect.\,\ref{subsec:examplefit}.

For the Be stars, we more frequently exclude certain regions from the fit as their spectra often show strong emission, mainly in the Balmer lines, but also sometimes in \ion{He}{i} and other spectral lines, which cannot be used to estimate the photospheric parameters of the stars. For stars that show only moderate contamination by emission lines in some of the diagnostic lines, we follow the same procedure as for B-type stars. We note that their derived parameters in general have larger uncertainties. In particular, when excluding the often contaminated H$\beta$ line, the surface gravity is not as well constrained (see Sect.\,\ref{subsec:fit_combine}). For thirteen Be stars with strong emission lines, it is not possible to derive parameters based on the spectra and we estimate only rotational velocities as described in Sect.\,\ref{sec:vsini_fwhm}.

\subsection{Photometric fit}\label{subsec:photfit}
In addition to the constraints from the MUSE spectra, we make use of the high-quality HST magnitudes available for almost all stars (a handful of stars fall in the CCD gap or are too close to a saturated star). HST observations were obtained in three broad-band filters covering the SED from the near-UV to the red edge of the optical range, namely F225W, F336W, and F814W \citep[centered at 2250\,\AA, 3360\,\AA, and 8140\,\AA, respectively,][]{Milone2018}. The catalogue also includes errors for F225W and F336W magnitudes, computed as random mean scatter (RMS) of the several independent photometric observations that are available. No errors are given for the F814W filter as there was only one epoch of observations. After conversion to fluxes, and correcting the errors for small sample statistics, we estimate the F814W error assuming the relative flux error is the same as the relative error on the F225W flux measurement.

Similarly to the spectroscopic fit, we again find the best-fitting model based on a grid search. For each model in the \textsc{tlusty} grid of \Teff\, and \logg, which provides the Eddington flux at the stellar surface, we compute the observed flux by scaling it with the distance of the SMC of 60\,kpc \citep[e.g.,][]{Deb2010} and apply interstellar extinction using the python package \textsc{dust\_extinction} assuming E(B--V)=0.08 \citep{Graczyk2014}, a constant R$_\mathrm{V}$=3.1, and the reddening law of \citet{Fitzpatrick2019}. The stellar radius is one of the fitting parameters and we further scale the model flux by varying it between 2 and 20 \Rsun\, in steps of 0.1 \Rsun. For each model in this now three-dimensional grid, we then compute the flux of the scaled and reddened model by convolving it with the corresponding HST filter\footnote{Filter transmission curves as well as flux zero points for each filter are obtained from \url{http://svo2.cab.inta-csic.es/svo/theory/fps/index.php}.}. 

We refrain from letting the extinction and distance vary as this would lead to a higher degeneracy in the fit. Given the generally low extinction, the choice of E(B--V) has a small impact on the overall results, especially when combined with the spectroscopic fit.  The distance to the SMC is well-known, and small-scale variations in the distance (for example the position of a star within the cluster), do not play a role given the overall large distance to the SMC. We do, however, note that only three photometric datapoints are available and fixing the distance and extinction might lead to systematic errors.

We then compute a photometric $\chi^2_\mathrm{phot}$ for each model in the 3D-grid by comparing the three observed fluxes with the computed model fluxes according to:
\begin{equation}
    \chi^2_\mathrm{phot} =  \sum_{i=1}^n \frac{(\mathrm{obs}_i - \mathrm{model}_i)^2}{\sigma_i^2} \quad,
\end{equation}
where obs$_i$ are the measured fluxes in the before-mentioned three HST filters, model$_i$ the filter-convolved model fluxes in the same filters, and $\sigma_i$ corresponds to the individual errors of the three flux measurements (i.e., n=3 for most stars). 

For a handful of stars, no photometric fit is possible as they have no HST magnitude available. Similarly to the spectroscopic fit, the photometric fit of Be stars is complicated by the circumstellar disk, in particular by excess flux in the near-IR \citep[see e.g.,][]{Rivinius2003}. While the F225W and F336W magnitudes should be relatively clean, the F814W magnitude is potentially affected by disk emission. As photometry and spectroscopy are not taken simultaneously, it is possible that the contamination varies between the two because of the variability of the disk. Given that we combine the photometric fit with the spectroscopic one, we generally find an acceptable fit for the Be stars, but with larger uncertainties than for B-type stars. We refrain from neglecting the F814W-magnitude for Be stars as we would be left with only two observed data points (i.e., F225W and F336W).

\subsection{Combination of the two fits}\label{subsec:fit_combine}
We combine the two separate fits by adding up the computed $\chi^2$ for each model with the same \Teff\, and \logg, but different \vsini\, and radius:
\begin{equation}
    \chi^2_\mathrm{global} = \chi^2_\mathrm{spec} + \chi^2_\mathrm{phot} \quad .
\end{equation}
By doing this without a further normalization, more weight is given to the spectroscopic fit in comparison to the photometric fit when more spectral lines (and thus more data points) are available. For spectra with a lower S/N or highly contaminated spectral where only few lines are used, the photometric fit is weighted more strongly. For most stars, both individual fits agreed within the errors, with the exception of some of the SB1 systems. A handful of additional stars were flagged based on such a discrepancy as it might be indicative of undetected binaries. 

The best-fitting model in this four-dimensional grid in terms of \Teff, \logg, \vsini\, and stellar radius is found by locating the minimum in the obtained $\chi^2_\mathrm{global}$ distribution for each parameter. We rescale the obtained $\chi^2$ distribution such that the model with the minimum $\chi^2$ has a $\chi^2_\mathrm{global}$ equal to the degrees of freedom in the computation. We furthermore give 2-$\sigma$ errors based on estimating the 95\%-confidence interval on the parameter range after rescaling the $\chi^2$. These are formal errors and do not take into account systematics, which could arise, for example, from fixing the microturbulent velocity. The limited availability of diagnostic lines in the observed spectra and the limited available model grid can affect the final parameters even beyond the give uncertainty ranges. Although the formal fitting errors in the derived stellar properties listed are quite small, the actual errors in the physical properties may be considerably larger because of systematic effects.

Typical uncertainties on \Teff, \logg, and \vsini\, are of the order of 1000\,K, 0.25\,dex, and 70\,\kms, respectively. While the \vsini\, uncertainty is in principle higher for higher rotational velocities, the dominating factor is the S/N ratio in the spectra. The errors derived for Be stars are, on average, higher than for B-type stars as fewer spectral lines can be used on average, and the F814W-magnitude might be impacted by the disk. In particular, this affects the determination of the surface gravity as for Be stars the Balmer lines (mainly also H$\beta$, which is \logg-sensitive and used as diagnostic line for B-type stars) is contaminated by emission. As mentioned above, when using several spectral lines, the observed rotational velocity has a better accuracy than the formal velocity resolution of MUSE. 

For $\sim$30 stars, the derived effective temperature reached the edge of the \textsc{tlusty} grid, when the associated errors were included. For those stars, we computed additional \textsc{tlusty} models extending to \Teff=10000\,K (L. Mahy, priv. comm., covering the same parameter range in \logg\, and \vsini\, as the nominal models) and repeated the analysis. If the error bars still hit the edge of the grid, we only provide upper or lower limits. This occurred predominantly for Be stars with strong emission, for which the stellar parameters are difficult to constrain. Due to the infilling of the Balmer lines, in particular a diagnostic for the surface gravity is lacking, which propagates to the determination of the effective temperature.
Best-fit parameters with associated uncertainties are given in Table~\ref{tab:all_params}. Diagnostic plots for each star (similar to the examples given in Figs. \ref{Fig:specfit_id654}, \ref{Fig:photfit_id654}, and \ref{fig:chi2_combined_654}) are available at the CDS.

\subsection{A test case: star \#654}\label{subsec:examplefit}

We provide a proof of concept for one star, \#654, and use it to test the assumptions on \vmac, distance and extinction. \#654 is a B-type star with a F814W magnitude of 16.3\,mag and a spectral type of B2 \citepalias{Bodensteiner2020a} located in the turnoff (TO) region of NGC 330. \#~654, also referred to as NGC\,330-095, was classified as B3\,III by \citet{Evans2006}  and its stellar parameters were further investigated by \citet{Hunter2008a}. Given that we found no significant RV variability \citepalias{Bodensteiner2021} we presume it is a single star \citep[which agrees with the classification in][]{Hunter2008a}.

\begin{figure} \centering
    \includegraphics[width=0.99\hsize]{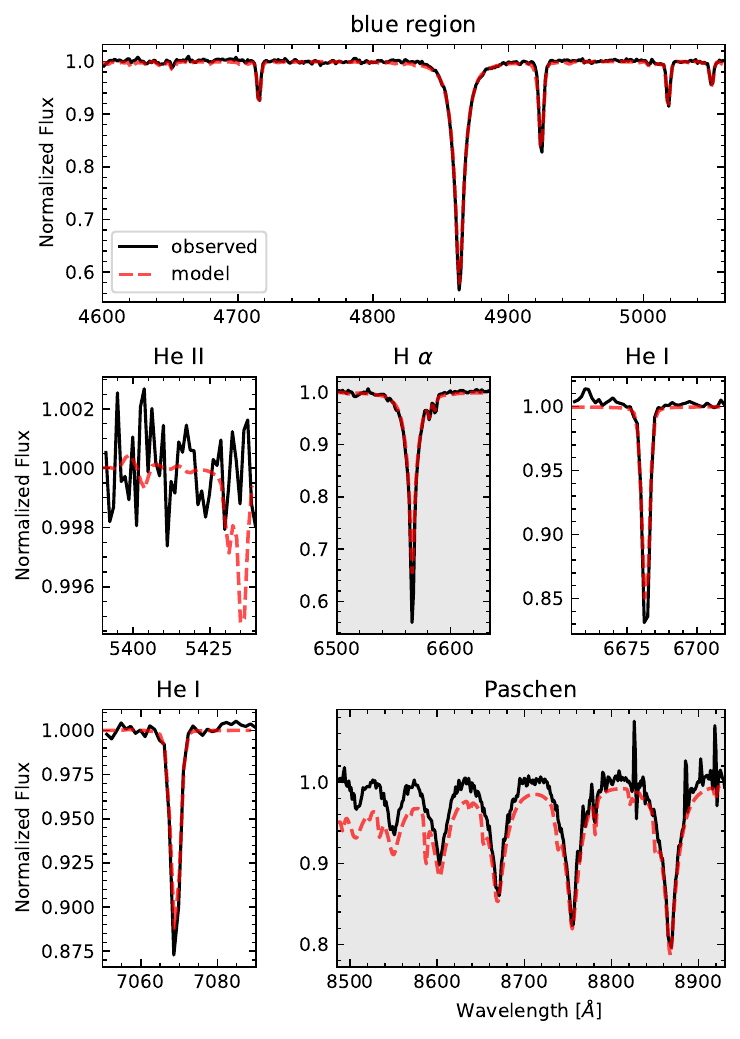}
    \caption{\label{Fig:specfit_id654} Spectroscopic fit for star \#~654. Each panel, one for each of the six diagnostic spectral regions, shows the combined spectrum (black) and the best-fitting \textsc{tlusty} model (red). Grayed-out regions are only shown for comparison but were not included in the fit.}
\end{figure}

\begin{figure} \centering
    \includegraphics[width=0.99\hsize]{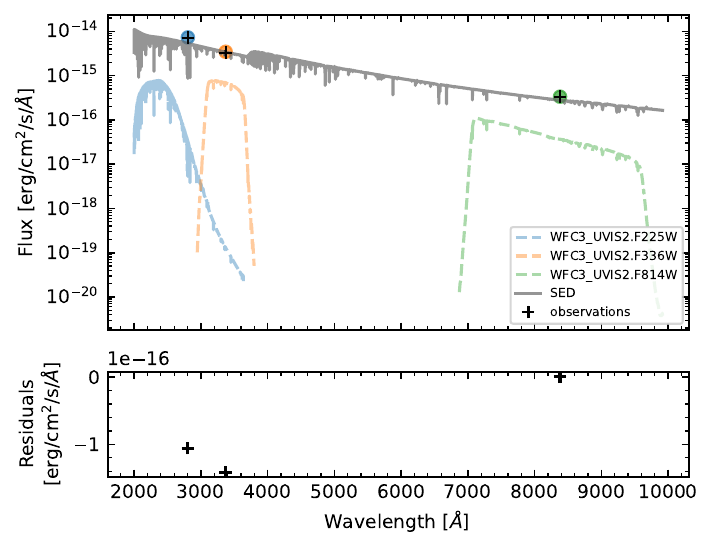}    \caption{\label{Fig:photfit_id654} Photometric fit for star \#~654. The top panel shows a comparison between the observed HST fluxes (black crosses) and the flux (colored circles) computed from the best-fit model (gray line) by convolution with the HST filters (colored lines). The bottom panel gives residuals.}
\end{figure}

\begin{figure} \centering
    \includegraphics[width=0.99\hsize]{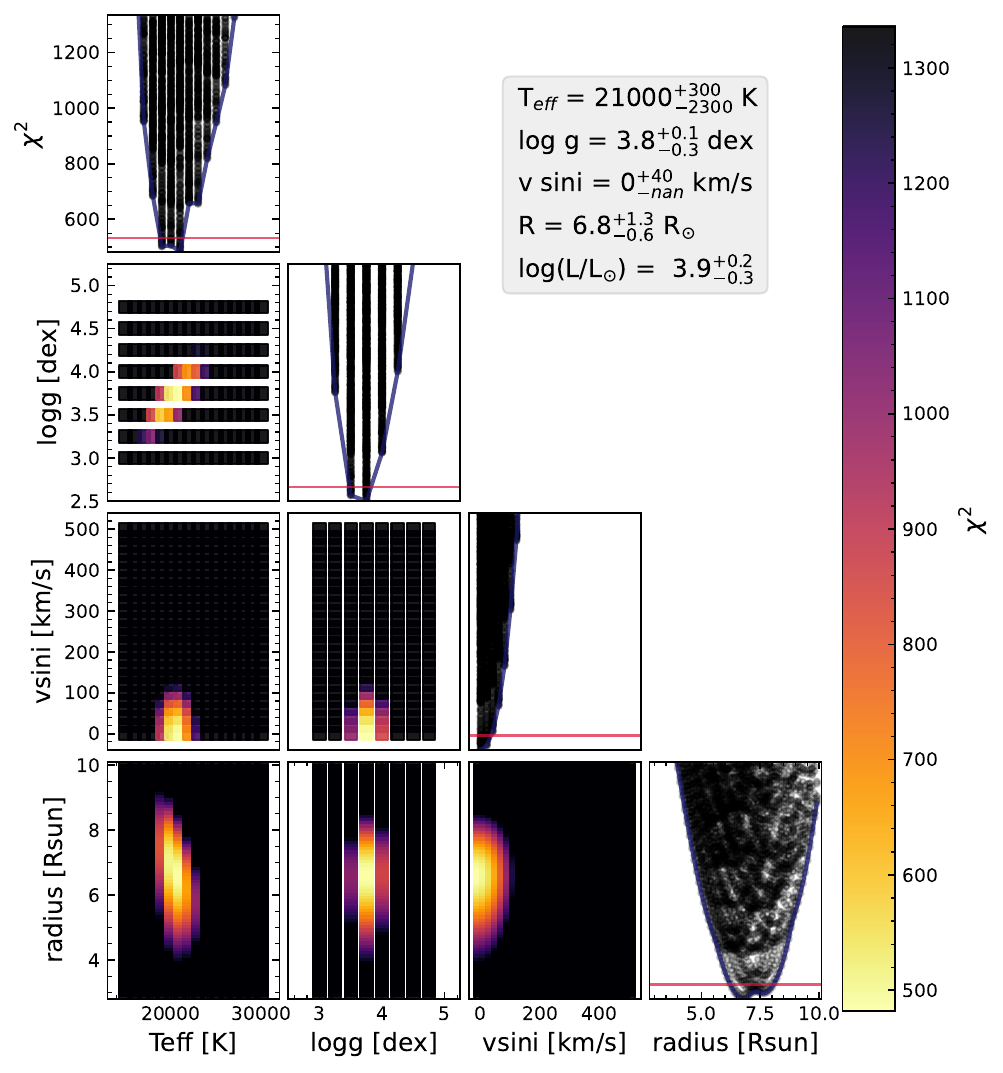}
    \caption{\label{fig:chi2_combined_654} Combined fit for star \#~654. The diagonal gives the $\chi^2$-distribution as a function of each parameter and the panels below show 2-dimensional $\chi^2$-maps. The red line marks the 95\% confidence level.}
\end{figure}

The spectroscopic fit of the observed MUSE spectrum, shown in Fig.\,\ref{Fig:specfit_id654}, is sensitive to the effective temperature, the surface gravity and the projected rotational velocity. The photometric fit, shown in Fig.\,\ref{Fig:photfit_id654}, constrains mainly the effective temperature and the radius. The global $\chi^2$ distributions derived of the combined fit can be seen in Fig.\,\ref{fig:chi2_combined_654}.

We find that \#\,654 has an effective temperature of $21000^{+300}_{-2300}$\,K a surface gravity of $3.8^{+0.1}_{-0.3}$\,dex, and a projected rotational velocity below 40\,\kms. Given the estimated radius of $6.8^{+1.3}_{-0.6}$\,\Rsun, the luminosity of the star is 
$\log$(L/\Lsun) = $3.9^{+0.2}_{-0.3}$.
The effective temperature is in good agreement with the one expected for a spectral type B2. The surface gravity agrees with the position of \#654 at the cluster TO, implying that it is close to the terminal-age MS (TAMS), as well as with the B3\,III classification of \citet{Evans2006}. The obtained radius is slightly higher than the typical radius of a B2 star on the MS, but probably in agreement with the radius of such a star at the TAMS. The rotational velocity of the star can be constrained below the formal resolution limit of MUSE because of the high S/N of the spectra and the large number of spectral lines used in the fit. 

\citet{Hunter2008a} assumed the effective temperature \Teff\,=\,18450\,K from the spectral type, and based on this inferred \logg\,=\,3.45, and \vsini\,=\,20\,\kms\, from a comparison to \textsc{tlusty} models. Assuming E(B--V) = 0.06, R$_\mathrm{V}$\,=\,2.72, the reddening law of \citet{Bouchet1985}, and bolometric corrections from \citet{Balona1994}, they derived $\log(\mathrm{L}/$\Lsun)\,=\,3.67. While the rotational velocity is in good agreement, the known degeneracy between \Teff\, and \logg\, (also illustrated in Fig.\,\ref{fig:chi2_combined_654}) explains the lower derived \logg\, by \citet[][, i.e., decreasing \Teff\, by 1000\,K decreases the \logg-estimate from the Balmer lines by approximately 0.1\,dex, which brings the estimate in good agreement]{Hunter2008a}. Despite the different assumptions on reddening and the different techniques used, the luminosities are comparable.

The only other star in common with the samples of \citet{Evans2006} and \citet{Hunter2008a} is \#\,57 or NGC\,330-036. It was classified as B2\,II star by \citet{Evans2006}, and \citet{Hunter2008a} report \Teff\,=\,21200\,K (again inferred from the spectral type), \logg\,=\,3.25\,dex, \vsini\,=\,39\,\kms, and $\log$(L/\Lsun)\,=\,4.33. In our approach, we find comparable parameters, namely \Teff\,=\,$24000^{+700}_{-200}$\,K, \logg\,=\,$3.5^{+0.1}_{-0.1}$\,dex, \vsini\,$<$\,50\,\kms, R\,=\,$10.3^{+0.1}_{-0.1}$\,\Rsun, and 
$\log$(L/\Lsun)\,=\,$4.5^{+0.1}_{-0.1}$. Again, the lower assumed effective temperature in \citet{Hunter2008a} will lead to the lower \logg\, derived. Overall, we thus find a good agreement between the studies, despite the lower resolving power of the MUSE spectra. To further test our method, we fitted the published \textsc{FLAMES} spectra of \#654 and \#57 with a similar approach than the MUSE spectra, retrieving stellar parameters that agree within the errors.


We further use \#\,654 to test our fitting assumptions. As described in Sections\,\ref{subsec:specfit} and \ref{subsec:photfit}, we assume a constant E(B--V)\,=\,0.08, A$_\mathrm{V}$\,=\,3.1 and neglect \vmac. When changing A$_\mathrm{V}$ to a lower value, for example  A$_\mathrm{V}$\,=\,2.74 as reported for the SMC \citep{Gordon2003}, the \Teff\, and $R$ derived from the photometric fit remain the same within the error bars. Similarly, changing E(B--V) by up to $\pm$20\% does not strongly impact the results. A larger change in the extinction, for example caused by a local overdensity of material around specific sources, would lead to a significantly higher \Teff\, (i.e., when assuming twice the extinction, E(B--V)=0.16, the photometric fit implies \Teff\,=\,28000\,K). Such a high temperature is excluded by the spectroscopic fit and the combined fit, which for this star is weighted towards the spectroscopic fit, does not change significantly. 

Finally, we use \textsc{iacob-broad} \citep{Simon-Diaz2014}, which determines the rotational velocity of a star based on a combined Fourier transform and goodness-of-fit method taking into account a non-zero \vmac. We find \vsini$=36^{+46}_{-11}$\,\kms\, and \vmac$=81^{+23}_{-81}$\kms. This again shows that \vsini\, can be constrained better than its formal accuracy. Further, the derived \vsini\, agrees with the one estimated in our approach, and the derived \vmac{} is consistent with our assumption of \vmac=0\,\kms.

The test case \#654 demonstrates the working principle of the fitting approach used here. As \#654 is a particularly slowly rotating star, we show the fitting of a second test case, a rapidly rotating Be star, in Appendix \ref{app:test688}.

\subsection{Estimation of \vsini\, based on FWHM measurements}\label{sec:vsini_fwhm}
As described above, several Be stars (13 out of the 83) could not be fitted with the \textsc{tlusty} approach as some of their spectral lines are dominated by emission. We thus resort to another commonly used method for estimating their projected surface rotation velocity. It is based on measuring the full-width at half maximum (FWHM) of spectral lines \citep[see e.g.,][]{Slettebak1975, Abt2002, Strom2005, Braganca2012, Garmany2015}, which to the first order is linearly related to the projected rotational velocity. Lines that are typically used for this purpose in B-type stars are lines in the blue part of the spectrum \citep[mainly \ion{He}{i} at $\lambda\lambda$\,4388 and 4713, \ion{C}{ii} at $\lambda$\,4267, or \ion{O}{ii} at $\lambda$\,4366; see e.g.,][]{Abt2002}. These \ion{He}{i} lines were also used to estimate the rotational velocity of Be stars \citep{Steele1999}. As described in Sect.\,\ref{subsec:specfit}, the FWHM-method neglects the effect of additional broadening by micro- and macroturbulence.

In \citetalias{Bodensteiner2021}, we fitted Gaussian profiles to spectral lines (\ion{He}{i} $\lambda$ 4922, among others) in order to measure RVs for the B and Be stars in the sample. From all stars with both FWHM estimates and \vsini\, measurements from the fitting described in Sect.\,\ref{subsec:specfit}, we derive a FWHM-\vsini\, relation which we then apply to the thirteen Be stars that we could not fit with the \textsc{tlusty} approach. We here use the FWHM measurements of the \ion{He}{i} line at $\lambda$ 4922\,\AA{} and exclude stars from the fit whose error bars hit the boundary of our grid, that is they are consistent with zero or reach velocities higher than 500\,\kms.

\begin{figure} \centering
    \includegraphics[width=0.99\hsize]{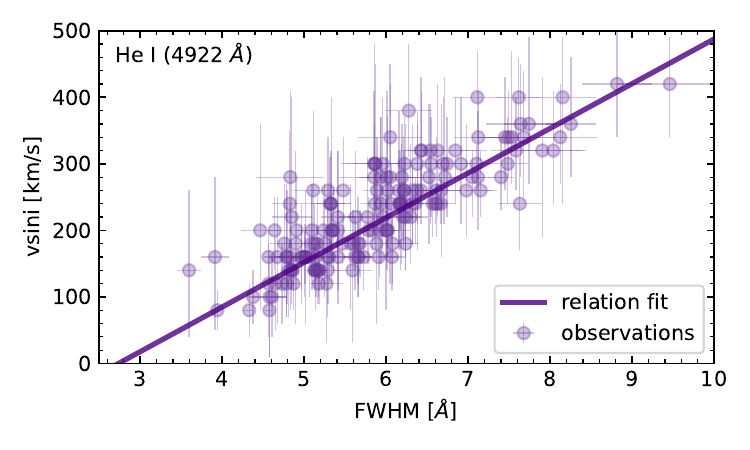}
    \caption{\label{Fig:corr_fwhm_vsini} Relation between the FWHM of the \ion{He}{i} line at 4922\AA{} with the \vsini\, estimated from the \textsc{tlusty} fit (purple line, see Eq.\,\ref{Eq:fwhm_vsini}). The purple points indicate the measured values for B-type stars .}
\end{figure}

The measured FWHM [\AA] and \vsini\,[\kms], shown in
Fig.\,\ref{Fig:corr_fwhm_vsini}, follow a roughly linear relation. Fitting it with a first-order polynomial, weighted by the measured errors in FWHM and \vsini\,, we find:
\begin{equation}\label{Eq:fwhm_vsini} 
    \varv\,\sin i\, [\mathrm{km}~\mathrm{s}^{-1}] = -183.3\,[\mathrm{km}~\mathrm{s}^{-1}] \\ + 67.1 \,[\frac{\mathrm{km}~\mathrm{s}^{-1}}{\AA}] \cdot \mathrm{FWHM} \,[\AA] \, ,
\end{equation}
with an RMS residual of 43\kms. As a consistency check, we repeat the fit for stars with fractional \vsini\, errors from the \textsc{tlusty} fit $<$0.2. The difference in the determined relation is negligible (as expected given the weighting by the errors). \citet{Ramirez-Agudelo2015} derived a similar relation for O-type stars using the \ion{He}{i} line at $\lambda$\,4922\,\AA. Their non-linear relation deviates from our linear relation for lower rotational velocities (below $\sim$150\,\kms). Otherwise, the obtained rotational velocities using the relation by \citet{Ramirez-Agudelo2015} agree with ours within the RMS.

\subsection{Deriving the intrinsic rotational velocity distribution}\label{subsec:deconvolve}

The sample sizes of the B- and Be-star in NGC\,330 are large enough to estimate their current rotational velocity distribution. Assuming that their rotation axes are randomly distributed, we can infer the probability density function, p(\ve), using the iterative procedure of \citet{Lucy1974} as implemented by \citet{Dufton2013}. This procedure has been used previously for B-star samples in the VFTS \citep{Evans2011} of 30 Doradus \citep{Dufton2013, Dufton2022}, and for NGC\,346 \citep[][where further details can be found]{Dufton2019}. Here, convolutions were undertaken separately for both the B- and Be-type samples in NGC\,330, as well as for the total sample. In all cases, the stars classified as apparently single and as SB1 candidates were included. The latter may not have a random distribution of orbital (and hence by implication rotational) axes of inclination, as discussed by \citet{Ramirez-Agudelo2015}. However, given the low number of SB1s, and combined with the likely presence of undetected binaries in the apparently single star sample, the overall distribution of rotational axes of the entire sample should be close to random. 

In addition, we have also re-derived rotational distribution functions for equivalent samples in the VFTS and NGC\,346 surveys. The \vsini-estimates for both the single and SB1 sub-samples were combined using the following sources: \citet{Dufton2013, Dufton2022} for the VFTS apparently single stars, \citet{Garland2017, Dufton2022} for the VFTS SB1 primaries, and \citet{Dufton2019} for the NGC\,346 targets. No correction has been made to the \vsini-estimates for gravity darkening \citep[see, for example,][]{Rivinius2013, Zorec2016}, in order to maintain consistency with our NGC\,330 estimates.

Care should be taken when interpreting the small-scale variation in the rotational distributions presented here. Even for the largest data sets, the implied stochastic uncertainties are significant and could lead to spurious small-scale structure. 
Nevertheless, we compare the intrinsic rotational velocity distribution to those derived for B and Be stars in 30 Doradus and NGC\,346 \citep{Dufton2013, Dufton2019, Dufton2022} in Sect.\,\ref{subsec:vsini_others}.\newline

\section{Stellar parameters of the massive star population}\label{sec:fitting_results}
We attempt a combined spectroscopic and photometric fit for 274 B and Be stars. However, the spectra were too noisy or contaminated, or photometric measurements were missing for 23 stars such that no stellar parameters could be determined. For a subset of the 23, namely thirteen stars, the projected rotational velocity could be derived based on the FWHM-\vsini-correlation described above. In total, this provides us with stellar parameters of 251 stars and \vsini-measurements of additional thirteen stars (264 stars in total), including presumably single stars and SB1s.

\subsection{Temperatures, surface gravities, radii and luminosities}\label{subsec:stellar_params} 

\begin{figure*} \centering
    \begin{minipage}[l]{0.49\textwidth}
        \includegraphics[width=0.98\textwidth]{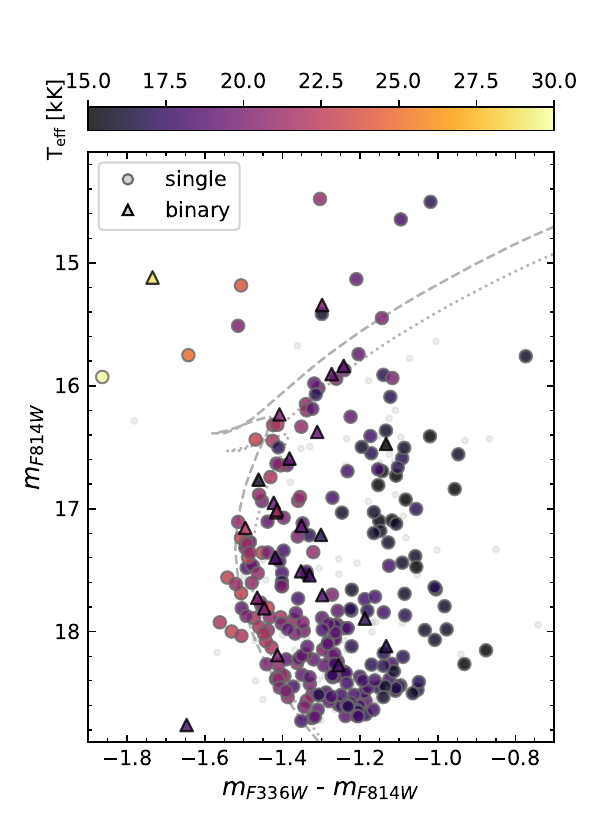}
    \end{minipage}
    \begin{minipage}[r]{0.49\textwidth}
        \includegraphics[width=0.98\textwidth]{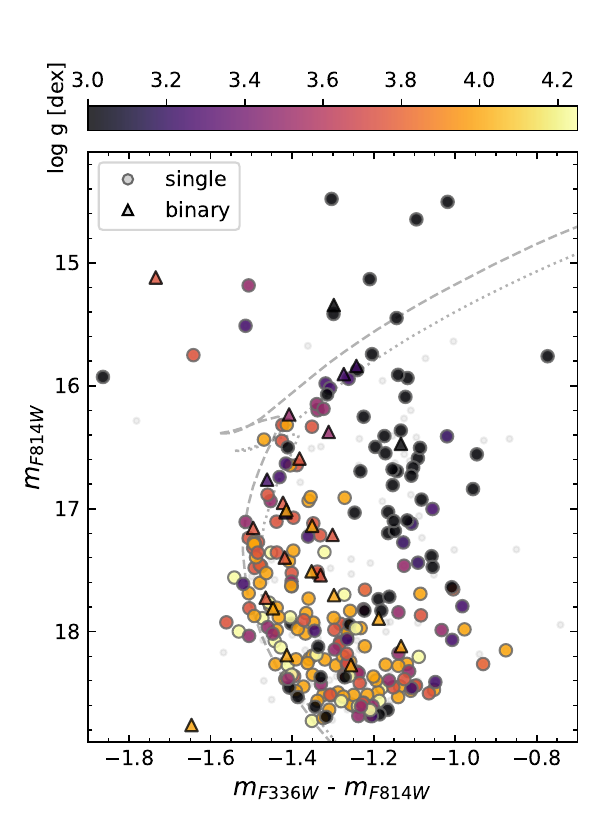}
    \end{minipage}
    \begin{minipage}[l]{0.49\textwidth}
        \includegraphics[width=0.98\textwidth]{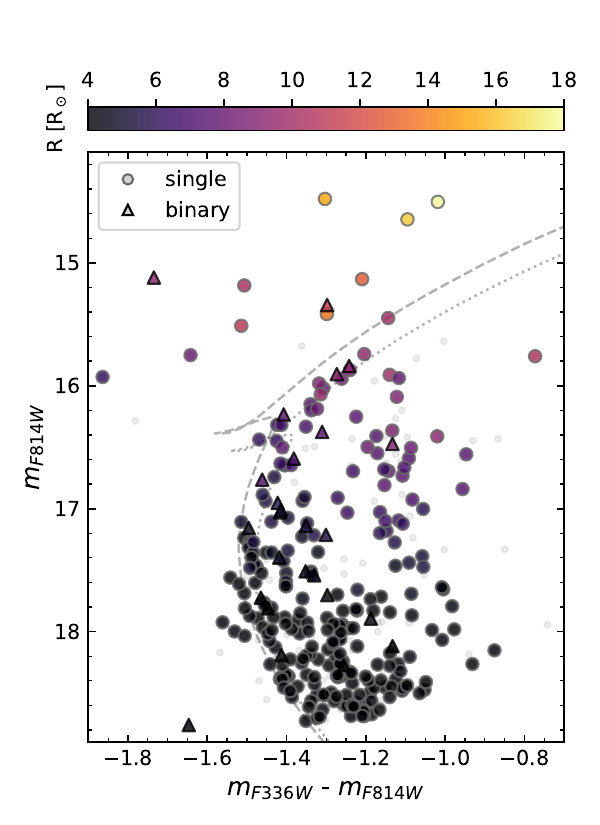}
    \end{minipage}
    \begin{minipage}[r]{0.49\textwidth}
        \includegraphics[width=0.98\textwidth]{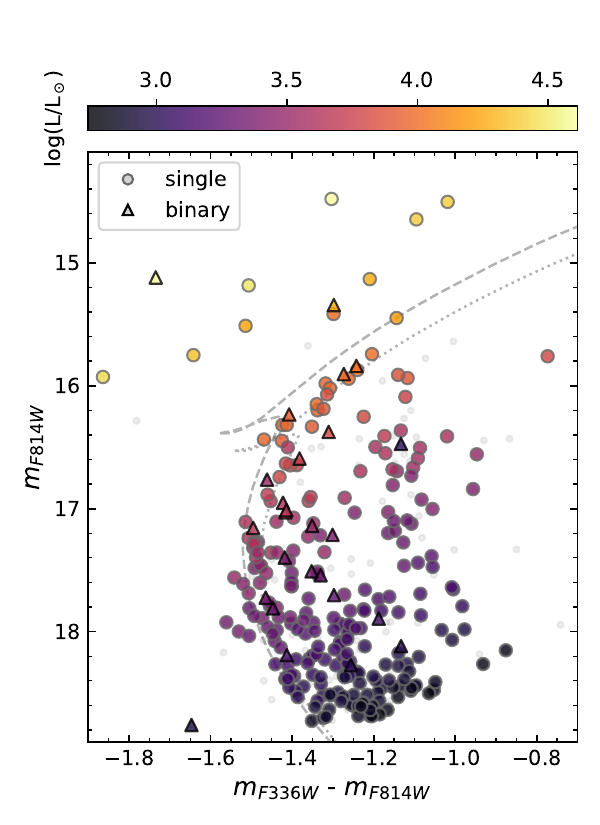}
    \end{minipage}
    \caption{\label{Fig:cmds_tlusty} CMDs of NGC~330 color-coded by the best-fitting physical parameters. Circles indicate presumably single stars while triangles indicate stars classified as SB1s. The isochrones are single-star non-rotating Padova isochrones of 35 (dashed line) and 40\,Myr (dotted line).}
\end{figure*}

Using the HST magnitudes of \citet{Milone2018}, we construct color-magnitude diagrams (CMDs) to investigate the distribution of the stellar parameters for the combined fit. Figure\,\ref{Fig:cmds_tlusty} shows that the temperature increases with the color ($m_\mathrm{F336W}-m_\mathrm{F814W}$) of the stars, as expected, but has a significant scatter. This is especially the case for the SB1s (see below) and for the stars above the cluster TO. The radius of the stars increases for the brighter stars, as well as for the more evolved stars, in line with the expectation that stars increase their radius at the end of the MS. The measured surface gravities generally also reflect this: they decrease for the more massive and more evolved stars. They are not corrected for the effect of gravity darkening. The estimated values show again a larger scatter, which is most likely due to the significant uncertainties in the \logg\, estimates.

In Fig.\,\ref{Fig:cmds_tlusty} we also indicate previously detected SB1s. While their derived radii and luminosities are in agreement with expectations, their temperatures and surface gravities show a larger scatter. The surface gravities of the SB1s do not show a trend in comparison to the single stars, but the effective temperatures of the SB1s seem, on average, slightly lower than the ones of single stars at the same locations in the CMD. The latter is consistent with spectral contamination from a presumably cooler unseen secondary. The greater scatter in the gravity estimates would also be consistent with such contamination by an unseen companion in, for example, the gravity-sensitive wings of Balmer lines.

One particularly hot star stands out in the CMD at m$_\mathrm{F814W}$=15.9. This star, \#\,420, is one of the two stars showing \ion{He}{ii} absorption in its spectrum. The spectrum is furthermore dominated by emission lines, hampering the quality of the fit. As shown in Fig.\,\ref{Fig:cmds_tlusty}, the \logg\, estimated for \#\,420 is low given its position in the CMD. This is probably connected to the fact that the estimated effective temperature of 30\,000\,K is at the edge of the used \textsc{tlusty} B-star grid, and the star is probably hotter (which would imply a higher \logg).

\subsection{Projected rotational velocities of B and Be stars}\label{subsec:vsini_dist}
\begin{figure} \centering
    \includegraphics[width=0.99\hsize]{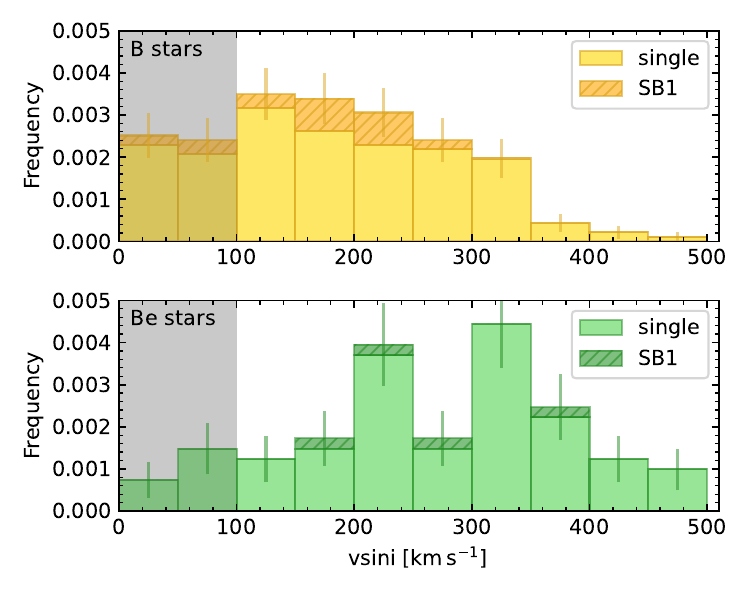}
    \caption{\label{Fig:vsini_hist_binary} Stacked histogram of the projected rotational velocities of presumably single stars and SB1s. The top panel shows the B-type stars (single stars in yellow, SB1s in orange) and the bottom panel shows Be stars (single stars in light green, SB1s in dark green). The gray shaded region again corresponds to the formal MUSE sensitivity limit while the indicated errors are counting errors.}
\end{figure}

As mentioned before, information about the projected rotational velocity is available for 264 stars. Figure\,\ref{Fig:vsini_hist_binary} shows that the \vsini\, of most B-type stars is around 100-250 \kms\, with a tail to high velocities up to 500\,\kms. The projected rotational velocities of Be stars show a distribution that is shifted to higher velocities: the majority of the Be stars are rotating with inferred projected velocities between 200-400\,\kms. 

The median projected rotational velocity of the B-type stars is around 170\,\kms\, while it is at $\sim$250\,\kms\, for the Be stars. We perform a Kuiper test and reject the null-hypothesis that both are drawn from the same parent distribution at the $10^{-3}$-significance level. The observed projected rotational velocity distributions of B and Be stars are therefore statistically different from each other, in agreement with findings from previous works \citep[see e.g.,][]{Hunter2008a, Dufton2022}.

In Fig.\,\ref{Fig:vsini_hist_binary} we also compare the projected rotational velocities of B and Be stars classified as presumably single or SB1s. Given the low observed binary fraction in NGC~330, there are only a handful of SB1s in the sample. Despite the low number statistics (or because of it), we find no obvious differences in the projected rotational velocity distributions of SB1s relative to presumably single stars. There is one rapidly rotating Be star classified as SB1 (with \vsini\,>350\,\kms), which might be a mass gainer in previous binary interactions with an undetected companion.  

We do, however, note that there are no B-type SB1s rotating at velocities $>300$\,\kms. This is similar to the findings of \citet{Ramirez-Agudelo2015}, who studied the primaries of O-type binaries in 30\,Doradus and found a lack of stars rotating with \vsini\,$>$\,300\,\kms, in comparison to single O-type stars. They interpret this as indication that the most rapidly rotating single O stars are actually binary interaction products \citep{deMink2013}. While their sample is dominated by SB1s, primaries and secondaries in SB2s were identified based on the luminosity ratio, which might be misleading in the case of Algol systems \citep[see e.g.,][]{Mahy2020a}. Investigating Galactic O-type stars, \citet{Holgado2022} found that the projected rotational velocity distributions of single stars and SB1s are different, and that the most rapid rotators (\vsini\,>\,300\,\kms) appear as single stars.

Another possible reason for the lack of very rapidly rotating SB1s could be an observational bias, with the more rapid rotators having broader spectral lines. In turn this leads to larger uncertainties in the RV estimates with only targets with large RV variations being identified. Further investigation of this possible observational bias is needed to confidently answer whether there are indeed no rapidly rotating SB1s in NGC~330.

The CMD colored by the derived \vsini\, values in Fig.\,\ref{Fig:cmd_vsini} shows a similar trend as Fig.\,\ref{Fig:vsini_hist_binary}: the stars populating the Be star region rotate, on average, more rapidly than stars on the MS. Furthermore, the lower MS (m$_\mathrm{F814W} > 17.8$\,mag) seems to show a trend between projected rotational velocity and color: while stars at the blue part of the MS rotate on average slower, the stars at the redder part of the MS rotate more rapidly. In particular, there are also some rapidly rotating B and Be stars at a color of around $-1.2$\,mag and m$_\mathrm{F814W}\,\sim18.5$\,mag. Those fainter stars usually have a lower S/N in their spectra, introducing a larger error in the derived rotational velocities. 

The variation of the \vsini\, along the MS band might indicate differences in the intrinsic rotational velocities of MS stars \citep[as predicted by theory, see e.g.,][and Sect.\,\ref{sec:NGC1818}]{WangC2022}. Given that stars on the lower MS are generally unevolved, most of them are not spun-up mass gainers in post-interaction systems. However, some of them  could be undetected binaries in which the spectral lines of the components are not fully deblended and thus mimic a larger observed rotational velocity. 

Figure\,\ref{Fig:cmd_vsini} furthermore shows that most single B-type stars above the cluster TO are slow rotators. Additionally, there are both slowly and moderately rotating Be stars above the cluster TO, which are both classified as single and SB1s. We will discuss the projected rotational velocities of stars in different regions further in Sect.\,\ref{subsec:vsini_regions}.

\subsection{The Hertzsprung-Russell diagram}
Despite the relatively large uncertainties in the obtained parameters, we use the effective temperatures and luminosities to construct an HRD (see Fig.\,\ref{Fig:gridsearch_HRD}). Overplotted are single-star evolutionary tracks and isochrones of 35 and 40\,Myrs from \citet{Georgy2013} assuming an initial rotational velocity of 50\% critical. Given that the cluster is expected to contain a significant number of binary interaction products, they are mostly to guide the eye.

Given the large uncertainties on the individual parameters, in particular \logg, we refrain from computing spectroscopic mass ($M_\mathrm{spec} \propto g / R^2$, so our typical \logg-uncertainties of $\pm0.25$\,dex translate into a factor of 2 in mass). A comparison of the star's position with respect to the isochrones allows to roughly estimate their evolutionary mass.

\begin{figure} \centering
    \includegraphics[width=0.99\hsize]{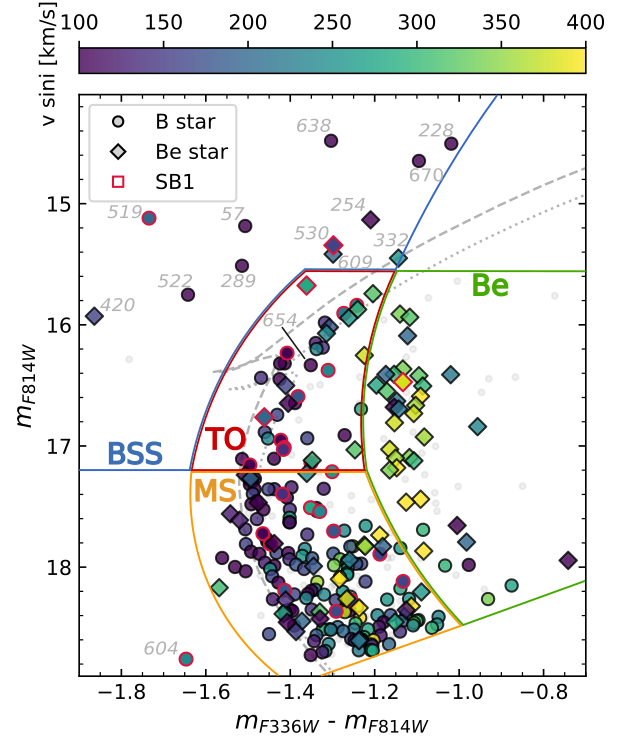}
    \caption{\label{Fig:cmd_vsini} CMD of NGC~330 constructed from HST photometry \citep{Milone2018}, color-coded by \vsini. For a reference, we provide two Padova isochrones at 35 (dashed line) and 40\,Myr (dotted line). B-type stars are indicated by circles while Be stars are indicated by diamonds. The color bar only includes \vsini\, values between 100\,\kms\, and 400\,\kms\, such that the few stars that rotate more rapidly fall in the last bin. Detected SB1s are indicated by red frames. The example star \#\,654, peculiar stars and stars above the TO are marked by their ID. The four regions defined in \citetalias{Bodensteiner2021} are also indicated.}
\end{figure}

\begin{figure} \centering
    \includegraphics[width=0.99\hsize]{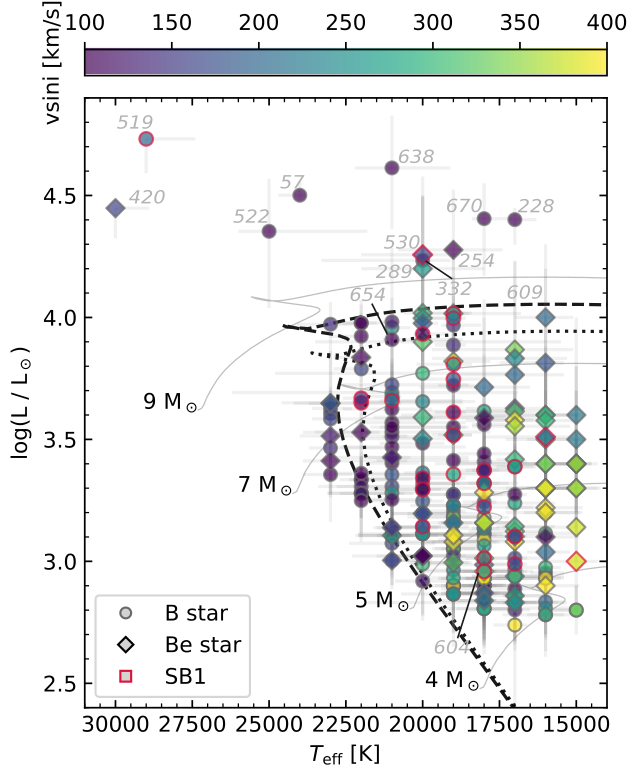}
    \caption{\label{Fig:gridsearch_HRD} HRD of B and Be stars in NGC~330. Presumably single B and Be stars are marked by a circle and diamond, respectively, while detected SB1s are indicated by symbols with red frames. The projected rotational velocity of is given by the color of the symbol. To guide the eye, we overplot evolutionary tracks from \citet{Georgy2013} assuming Z=0.002, an initial rotation rate of 50\% critical, and for different initial masses as indicated in the plot. We further overplot 35\,Myr (dashed) and 40\,Myr (dotted) isochrones with the same assumptions. Stars above the TO in the CMD and peculiar stars are marked by their ID. The limited availability of diagnostic spectral lines, photometric data points, and suitable models can lead to actual errors larger than the formal fitting errors indicated here.}
\end{figure}

In the HRD, we indicate the IDs of stars located above the cluster TO in the CMD (Fig.\,\ref{Fig:cmd_vsini}). This demonstrates that stars situated above the TO in the CMD are also above the TO in the HRD. In particular, the slowly-rotating single B-type stars above the TO (namely \#\,522, 289, 57, 638, 670, and 228), which are above the TO in the CMD and seem to follow a younger isochrone, also trace a similar, younger isochrone in the HRD. In contrast, the particularly faint and blue star \#\,604 does not show peculiar stellar parameters. This is in line with the possible explanation we put forward in \citetalias{Bodensteiner2020a} where we proposed that the visible star might have a stripped companion, which contributes to the colors but not to the spectral lines.

The evenly-spaced values of \Teff\, reflect the steps in the \textsc{tlusty} grid used.
The apparent lack of stars towards the isochrones at lower luminosities (i.e., $\log$\,$L$/\Lsun) < 2.8) is probably mainly due to our selection criterion in magnitude, which preferably selects cooler stars at the same luminosity. At higher luminosities, several stars are to the left of the two isochrones (although consistent with the 35\,Myr-isochrone within the errors). This might imply that rejuvenation took place. It could also be due to the choice of 50\%-critical initial rotation in the isochrone and might imply that this is not a valid assumption for all the stars. 

Several stars, also of lower mass, populate the post-MS region where they should not be situated according to standard single-star stellar evolution. While some of them could be binary-interaction products, the large errors in derived effective temperatures (i.e., $\sim$1000\,K) and the uncertainties in the evolutionary tracks such as the adopted rotational velocity and overshooting most likely blur the picture. Undetected binary systems might lead to a further broadening of the observed MS given that the dilution by a companion star might light to a lower derived temperature while the derived luminosity is higher than for a single star. Therefore, and given the larger error bars, we focus on the CMD in the following, which benefits from the high accuracy of HST photometry and provides a clear overview of the cluster.

The HRD indicates that a majority of the rapidly rotating Be stars are towards the cooler end of the MS band. While the errors are larger for the Be stars, this may be due to a combination of two factors: Be star with a high projected rotational velocity are likely viewed equator-on. Due to the deformation of rapidly rotating stars, they have lower temperatures around their equator than around their poles \citep[see e.g., ][]{VonZeipel1924}. Additionally, in the cases where the photometry fit is weighed more strongly, the lower derived temperature might reflect the impact of the circumstellar decretion disk for Be stars \citep{Townsend2004}. As noted previously noted in \citetalias{Bodensteiner2020a} and by \citet{Hastings2021}, a large number of Be stars are unevolved and far from the cluster TO, implying they are binary interaction products.

\subsection{Rotational velocities for stars in different CMD regions}\label{subsec:vsini_regions}
In \citetalias{Bodensteiner2021} we defined different regions in the CMD to investigate possible differences in the observed and bias-corrected binary fractions (indicated in Fig.\,\ref{Fig:cmd_vsini}). We found that the observed and bias-corrected spectroscopic binary fractions differ significantly in the different regions. Here, we investigate if the \vsini-distributions also vary from region to region.

The regions, namely the main sequence (MS), turn-off (TO), blue straggler star (BSS), and the Be star region, are solely defined based on the position of stars in the CMD in comparison to the 40-Myr Padova isochrone (see Fig.\,\ref{Fig:cmd_vsini}). The association of stars to a region does not take into account their spectral appearance. In particular, the Be star region is defined as a region offset redward of the MS, not based on the presence of emission lines in the spectra. Several spectroscopically classified Be stars fall out of it, while five faint B-type stars fall in the Be star region. 

We note that this current division in regions considers the MS as one region, without further subdividing it in terms of a blue and red MS  \citep[as done for example by][]{WangC2022, Kamann2023}, despite the fact that there might be a possible bi-modality in the \vsini\, of the faintest stars (see Sect.\,\ref{subsec:vsini_dist}). This region is particularly interesting to investigate unevolved stars and pre-interaction binaries. In a future paper of this series, we will consider additional, fainter stars and perform this detailed investigation of the rotational velocties of lower MS.

\begin{figure} \centering
    \includegraphics[width=0.99\hsize]{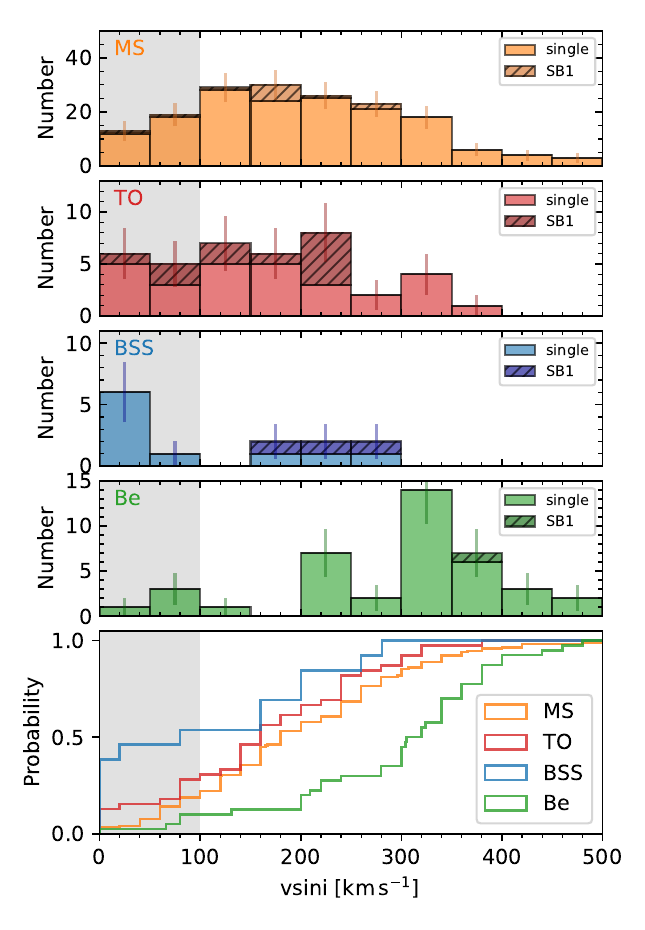}
    \caption{\label{Fig:regions_vsini} Projected rotational velocities of different stellar groups, defined based on the CMD (see Fig.\,\ref{Fig:cmd_vsini}). \textit{Top four panels:} From top to bottom, the panels show the \vsini\, distributions of star in the MS region, the TO region, the BSS region and the Be star region. In each panel, the detected SB1s are marked by a darker color, as indicated in the legend. \textit{Bottom panel:} Normalized cumulative distribution of the rotation rates of different groups, as indicated in the legend.}
\end{figure}

Figure\,\ref{Fig:regions_vsini} shows the velocity distribution of stars in the four regions. It demonstrates that the rotational velocities of MS and TO stars are rather similar. The projected rotational velocities in the BSS region are generally below 300\,\kms, which is lower than the stars in other regions. The stars in the Be star region show higher average velocities around \vsini$\sim 350$\,\kms. 

To verify if the observed differences are statistically significant, we perform a Kuiper test between each of the groups (see Tab.\,\ref{Tab:region_kuiper}). The difference in the rotational velocity distribution of Be stars compared to any other group is statistically significant (that is, the probability that the distributions are drawn from the same parent distribution is below 0.03\% for any of them). Despite the appearances, the \vsini-distributions of MS, TO, and BSSs are not significantly different from one another, which may in part be due to the low number of stars (i.e., only 13) in the BSS region. When considering only \vsini\, above 100\,\kms\, (which we adopt as formal MUSE sensitivity limit), we find similar results: the projected rotational velocities in the MS and TO regions are similar, while the Be star region is statistically different. The BSS region only contains six stars after this cut and can therefore not be used for such a statistical comparison.

\begin{table}\centering
    \caption{\label{Tab:region_kuiper} Kuiper-probability that observed \vsini\, distributions of different populations are drawn from the same parent distribution.}
    \begin{tabular}{lcccc} \hline \hline
            & MS & TO   & BSS  & Be      \\ \hline
        MS  & -  & 97\% & 12\% & <0.01\% \\
        TO  &    & -    & 75\% & <0.01\% \\
        BSS &    &      & -    & \quad 0.03\%  \\ \hline
    \end{tabular}
    \tablefoot{The populations are located in different regions in the CMD as indicated in see Fig.\,\ref{Fig:cmd_vsini}.}
\end{table}

Similar to the MS region, the TO region probably contains stars of different natures (e.g., truly single stars, pre-interaction binaries, or around the TO also currently interacting binaries). The observed rotation rates in these two regions are similar and show a large spread. There furthermore seem to be no significant differences between presumably single stars and SB1s. We do, however, note that there are a large number of SB1s with rotational velocities between 100 and 200 \kms\, in the TO region. These might coincide with a group of binary systems rotating at similar velocities detected by \citet{Ramirez-Agudelo2015}, who interpreted them as being spun up through tides.

The BSS region is populated by two different types of systems: On the one hand, there are apparently single B-type stars with particularly low rotational velocities. Those could be merger products, which are predicted to be slow rotators \citep[see e.g.,][]{Schneider2020}. On the other hand, there is a group of SB1 systems and emission line stars with \vsini$\sim250$\,\kms. Those could be currently interacting or post-interaction binary systems. 
In particular, the two SB1s with Balmer emission lines could be currently interacting binaries where the emission is formed in an accretion disk (possibly Algol-like systems in which the mass ratio was reversed). Some of the more slowly rotating SB1s might be binary systems in which tides may have led to a synchronization and therefore a spin-down.

Stars in the Be star region have a significantly different rotation rate distribution: on average, they are rotating more rapidly than stars in the other regions. This is in line with theoretical expectations: Be stars could be single stars approaching their critical rotation rate, or they could have gained their rapid rotation as mass gainers in previous binary interactions. The rapid rotation (and the additional IR emission from the circumstellar disk) leads to a redder color of the stars, populating the Be star region. We further note that there are four slowly rotating stars in the Be star region, which could appear as such because of inclination effects. Given the sample size of 47 targets in the Be star region, it seems, however, unlikely that four stars (i.e., almost 10\%), are viewed under low inclination angles.

\section{Comparison to previous observational studies}\label{sec:otherworks}
\subsection{Rotational velocities in the LMC cluster NGC~1818}\label{sec:NGC1818}

The cluster NGC~1818 is located in the LMC and has a similar age as NGC~330 \citep[$\sim$ 40\,Myrs,][]{Milone2018}. Similarly to NGC~330, NGC~1818 was reported to have an extended TO region. Additionally, NGC~1818 has a split MS (i.e., the main MS in NGC~1818 shows two components, in addition to the Be sequence), which is less pronounced in NGC~330. Given the lower metallicity in the SMC compared to the LMC, a comparison between NGC~330 and NGC~1818 might help assessing the impact of metallicity on the observed rotational velocity distributions.

Using VLT/FLAMES in GIRAFFE mode, 44 MS stars in NGC~1818 were studied spectroscopically and their rotational velocities were derived by comparing the observed \halpha- and \ion{He}{i}-line at $\lambda$6678 to synthetic spectra \citep{Marino2018}. All targets were selected to be on the MS (both the red and the blue MS) and span a brightness range m$_\mathrm{F814W}=15.5-19.0$.

The derived rotational velocities of the stars in NGC~1818 were found to vary as a function of the brightness and color of the star: stars attributed to the blue MS rotate, on average, slower than stars in the red MS. Be stars are found to be the fastest rotators, while stars around the TO are generally slow rotators \citep[see e.g., fig. 7 in][]{Marino2018}. This agrees with theoretical predictions that explain the split MSs observed in young star clusters with different rotational velocities \citep[e.g., ][]{Bastian2009, WangC2022}. A similar bi-model velocity distribution was also observed in the MS stars of the 100\,Myr-old LMC cluster NGC~1850 \citep{Kamann2023}.

While these results are based on only a low number of stars in NGC~1818 (e.g., four stars in the TO region, and four Be stars), they qualitatively agree with the results derived here for NGC~330, despite the different metallicity of the two clusters. Here, no difference was made between the red and blue MS. Distinguishing those might have an impact on the rotational velocity distribution of the lower MS. Indeed, the rotational velocities of stars in the MS box in Fig.\,\ref{Fig:cmd_vsini} appear to vary as a function of color: redder stars rotate, on average, faster.  A more detailed comparison of the rotational velocities in the different MS components will, however, be part of a subsequent paper in this series.

\subsection{Rotational velocities in 30 Doradus and NGC~346}\label{subsec:vsini_others}
\begin{figure} \centering
    \includegraphics[width=0.99\hsize]{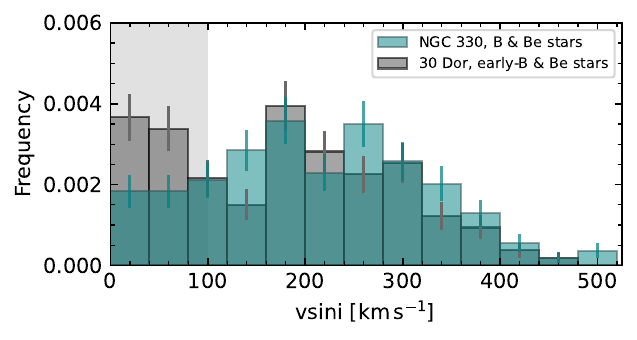}
    \caption{\label{Fig:comp_dufton} Normalized histogram of the projected rotational velocities of B and Be stars in NGC~330 (turquoise) compared to early-B and early-Be stars in 30 Doradus (gray) \citep{Dufton2013}, adopting the same bin size as in their fig. 3.}
\end{figure}

We compare our observed rotational velocities to the ones measured by \citet{Dufton2013} for early B-type stars in the 30\,Doradus region in the LMC in Fig.\,\ref{Fig:comp_dufton}, excluding stars classified as O9.5 (see their figure 3 for more details). We adopt a bin size of 40\,\kms, the same as chosen by \citet{Dufton2013}, and include all B and Be stars in NGC~330 \citepalias[including spectral types O9 to A0, see][]{Bodensteiner2020a}. We note that our spectral types are on average later than the ones studied by \citet{Dufton2013}, who focused on stars with spectral types B0-B3, implying that the stars studied have different masses. We refrain from restricting our sample to those early spectral types because it would reduce the sample size in NGC~330 to only $\sim$40 stars.

\citet{Dufton2013} reported a bi-modal \vsini\, distribution of B-type stars in 30\,Doradus, with $\sim25\%$ of the sample rotating slower than $\sim80$\,\kms\, and a second \vsini\, peak at around 200\,\kms. Focusing on the O stars in 30\,Doradus, \citet{Ramirez-Agudelo2013} found a similar peak at $\sim80$\,\kms, and a tail of high velocities up to 600\,\kms.
Our results indicate that there is no bi-modal \vsini\, distribution in NGC~330: the distribution is rather flat up to $\sim$300\,\kms\, after which it tails of. The apparent double-peaked feature around 200\,\kms\, is most likely due to the adopted binning (see also Fig.\,\ref{Fig:vsini_hist_binary}, where such feature is visible for the Be stars). Given the velocity-resolution of MUSE, however, a bi-modality cannot be ruled out. Indeed, it might be present in the lower MS (see Sect.\,\ref{subsec:vsini_dist}.)

Figure\,\ref{Fig:comp_dufton} furthermore indicates that the stars in NGC~330 rotate, on average, more rapidly than the stars in 30\,Doradus. We test the statistical significance of this finding by performing a Pearson $\chi^2$ test, which indicates that this result is statistically significant (the likelihood that both \vsini-distributions are drawn from the same parent distribution is $<10^{-3}$). When only considering B-type stars and excluding Be stars in NGC~330 (which on average show more rapid rotation rates) we still find a similar result, though with a lower confidence.

\begin{figure} \centering
    \includegraphics[width=0.99\hsize]{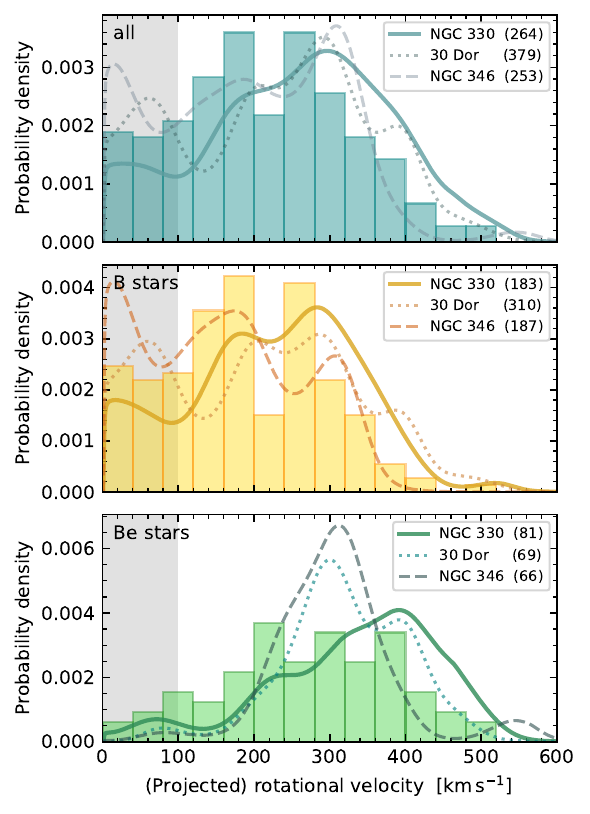}
    \caption{Projected and intrinsic rotational velocity distributions of B and Be stars in NGC~330. The normalized histogram in each panel shows the observed projected rotational velocities in NGC~330, while the solid line indicates the intrinsic distribution of rotational velocities, that is corrected for random inclination angles. The dotted and dashed lines show the intrinsic rotational velocity distributions for stars in 30 Doradus and NGC 346, respectively \citep{Dufton2013, Dufton2019, Dufton2022}. The number of stars in each sample is indicated in brackets. The \textit{top panel} shows B and Be stars combined, the \textit{middle panel} focuses on B-type stars, while the \textit{bottom panel} shows the Be stars only.}
    \label{Fig:deconvolved}
\end{figure}

We further compare the intrinsic rotational velocities of the B and Be stars in NGC~330 (see Sect.\,\ref{subsec:deconvolve}) with stars in 30 Doradus and NGC~346 (see Fig.\,\ref{Fig:deconvolved}). NGC~346 is a $\sim$2\,Myr-old, star-forming \ion{H}{ii} region in the SMC \citep{Dufton2013, Dufton2019, Dufton2022}. While NGC~346 is at a similar metallicity than NGC~330, the stellar populations in 30 Doradus and NGC~346 are significantly younger than the one in NGC~330. We include all stars classified in the three samples as single stars and SB1s and note that the different observing approaches have different binary detection probabilities. We again note that the different studies target a slightly different range of spectral types, or stellar masses: stars in NGC~330 have spectral types from B0 to B9 \citepalias[see][]{Bodensteiner2020a}, while the B-star sample in NGC~346 focuses on stars between B0 and B3, similar to the 30-Dor sample.

The intrinsic rotational velocity distribution of the entire NGC~330 sample shows no evidence for a significant bi-modality, but rather a peak at 300\,\kms\, (see Fig.\,\ref{Fig:deconvolved}). The comparison of B and Be stars demonstrates again that the Be stars rotate, on average, more rapidly than the B-type stars: their intrinsic distribution peaks around 400\,\kms, with only very few stars rotating slower than 150 \kms. The local maxima in the B- and Be-star distributions are most likely due to the binning of the rotational velocities and do not represent true features. 

In contrast, the overall sample of stars in 30\,Dor (and to a lesser extent in NGC~346) seem to have bimodal rotational velocity distributions. In general, there are fewer stars in NGC~330 with rotational velocities below 100\,\kms, and more rapidly rotating stars, in particular compared to NGC~346. The median rotational velocity of the stars in NGC~346 is significantly lower (i.e., $\sim$150\,\kms) than the one in 30~Doradus and NGC~330 ($\sim$230\,\kms\, in both clusters). 
Figure\,\ref{Fig:deconvolved} further indicates that the Be stars, which follow unimodal rotational velocitity distributions in all three populations, are more rapidly rotating in NGC~330 (their median rotational velocity is 350\,\kms) than the ones in the other two populations, and in particular than the ones in NGC~346 (with a median velocity of $\sim$300\,\kms).

\subsection{Discussion}
The comparison of the derived rotational velocities in the different observed samples is complicated by several aspects. Firstly, as mentioned previously, the mass ranges targeted by the different studies differ from each other. While the samples in NGC~346 and 30~Dor are comprised mainly of early-B type stars \citep[O9.5 to B3, corresponding to approximately 16 to 8 \Msun,][]{Dufton2013, Dufton2019}, the stars in targeted in NGC~330 and NGC~1818 are primarily of later spectral types (O9 to B9, corresponding to 16 to 3\,\Msun, with a peak at spectral type B5, see figure\,14 in \citetalias{Bodensteiner2020a}). The mass of a star impacts the strength of the stellar wind, which can spin-down stars if the mass-loss rates are high enough. However, as B-type stars show only weak winds, this should not lead to significant differences in the \vsini-distributions.

Another caveat in the comparison is that we consider populations of stars are at different metallicities. While NGC~330 and NGC~346 are both in the SMC (with a metallicity Z\,$\sim0.2\,\mathrm{Z}_{\odot}$), 30~Dor and NGC\,1818 are in the LMC (with $\mathrm{Z}\sim0.5\,\mathrm{Z}_{\odot}$). As discussed in Sect.\,\ref{Sec:intro5}, stars at lower metallicity should on average have higher projected rotational velocities because they are more compact and have weaker winds. The second effect, as mentioned above, is barely relevant for B-type stars. 
This is supported by the fact that we find generally higher rotational velocities for stars in NGC~330 compared to 30~Dor and NGC~346, while there is a general agreement with the stars in NGC~1818.

A third difference between the studies are the environments they target. As discussed in Sect.\,\ref{Sec:intro5}, Galactic studies found that cluster stars rotate on average faster than field stars \citep[e.g.,][]{Strom2005, Daflon2007, Garmany2015}. These differences were attributed either to differences in the star formation process \citep[cf.][]{Wolff2007} or to evolutionary effects \citep{Huang2008}. While the studies on NGC~330 and NGC~1818 focus on the central core of the clusters, the study of 30~Dor targets a more extended star-forming region. In contrast, stars in NGC~346 are mostly field stars \citep{Dufton2019}. The latter could be subject to sequential or continuous star formation. The lower \vsini\, in NGC~346 could at least in part be explained by the stars being predominantly field stars.

The last and main difference between the samples is their age: while NGC~346 and 30~Dor contain very young stars \citep[i.e., $<5-15\,$Myr,][]{Dufton2013}, the population of stars in NGC~330 and NGC~1818 are roughly 35-40\,Myr old ( which is also the reason for the later spectral types in NGC~330 and NGC~1818). Most stars and binary systems in NGC~346 and 30~Dor are unevolved, and the parameters detected there represent the initial parameters. In contrast, the more massive stars in NGC~1818 and NGC~330 have evolved off the MS. If in close-enough binary systems, those systems interacted, either by mass transfer or via a stellar merger. NGC~330 and NGC~1818 is therefore thought to be populated by such post-interaction products, either the spun-up mass gainers of mass transfer, or the (potentially slowly rotating) merger products \citep[e.g.,][]{deMink2013, Schneider2018}. In the case of mass transfer, their previously more massive companions could now be compact objects or might have exploded as supernova, and the systems are now detected as single stars. Due to rejuvenation, such systems are expected to be situated around or above the cluster TO. 

These expectations are in line with our observational findings in NGC~330: the rapidly rotating, mainly single Be stars could be such spun-up companion stars. The slowly rotating, apparently single stars above the cluster TO could be merger products. Finally, the moderately rotating stars above the cluster TO, among which some show emission lines and some are detected as binaries, could be currently interacting binary systems. While our observations and the theoretical predictions of stellar spins match qualitatively, other factors such as the metallicity of the stars might play an important role. A more quantitative comparison, which we will perform in a subsequent paper of this series, will allow to further disentangle the different aspects.

\section{Summary and conclusions}\label{sec:conclusions}
In this work, we have measured the stellar parameters of the presumably single stars and SB1s in NGC~330. For this purpose we used the available HST photometry as well as spectra that were combined from the six observed MUSE epochs. We determined stellar parameters by a two-component grid-search approach using the BSTAR2006 \textsc{tlusty} grid of model atmospheres.

The high quality of the HST observations allows us to constrain radii and luminosities. Given the availability of only three filters and the fixed extinction to a common value for all stars, systematic uncertainties may remain. The low-resolution spectra lead to a larger scatter (accompanied by larger relative uncertainties) in \Teff\, and \logg. Rotational velocities can be constrained even better than the formal MUSE resolution limit of around 100\,\kms\, when including several lines in the fit. In general, the derived stellar parameters agree with the expectations from the star's position in the CMD. Furthermore, we found no significant differences in the determination of the parameters of previously identified SB1 systems, apart from a larger scatter and larger uncertainties that could be due to a possible contamination of the observables by the unseen companion. 

Our observations show that a majority of B-type stars have \vsini$\approx100-250$ \kms\, with a tail to higher velocities. The projected rotational velocities of Be stars are on average higher and show a peak at around 200$-$400\,\kms. A Kuiper test demonstrates that the two distributions are statistically different. We further associated different regions in the CMD to different stellar populations: stars on the lower main sequence are associated to the MS region, stars around the turnoff to the TO region, blue straggler stars above the cluster turnoff to the TO region, and stars redward of the MS to the Be region.
We found that the projected rotational velocities of the stars in the Be region are significantly larger than the ones of stars on the MS, and in the TO and BSS regions. While BSSs seem to, on average, rotate more slowly than MS and TO stars, small sample sizes prevent us from confirming the significance of this result. We note that a subset of the BSS stars are apparently single, slowly rotating stars, which could be the products of stellar mergers.

Comparing the stars in NGC~330 in other stellar populations, we find an overall agreement with stars in NGC~1818, a LMC cluster of similar age. We further find that the stars in NGC~330 generally rotate more rapidly than O- and early-B-type stars in the young 30\,Doradus region in the LMC \citep{Ramirez-Agudelo2013, Dufton2013, Dufton2019}, and more rapidly than B-type stars towards the young SMC cluster NGC~346. Given that the latter two studies mostly targeted B-type MS stars, which in general have weak winds, spin-down due to stellar winds should not be able to explain the observed differences.

We propose that the higher rotation rates in NGC~330 could be rooted in the age difference between the clusters. In the young clusters (30 Dor and NGC\,346), the late-O and early-B stars will be relatively unevolved, whilst most binary systems will not have interacted. By contrast both of these phenomena may be important for the older cluster NGC\,330, where more massive stars had time to evolve and interact. Mass transfer in close binary systems might lead to a higher fraction of Be stars, which could be the mass gainers in previous binary interactions. The brightest Be stars close to the TO could, however, also be single stars with sufficiently high initial rotational velocities to spin-up towards the end of their MS evolution. The slowly-rotating, apparently single stars above the cluster TO could be merger products. This is in qualitative agreement with binary population synthesis predictions \citep[see e.g.,][]{WangC2020}.

In this work, we provided an extensive and homogeneous data set which can be directly compared to modern population synthesis models including the most recent treatment of binary interaction physics \citep[e.g.][]{Eldridge2009, WangC2020, WangC2021, Klencki2022}. Such a comparison will help to constrain currently implemented single- and binary-star evolution physics and allow a better understanding of massive star evolution. A possible future step to further investigate the properties of (at least the brightest) cluster members is to study their chemical abundances. In particular the BSS, which are excellent merger candidates, might show CNO-processed material on their surface such as nitrogen enhancement. Despite the low resolving power of MUSE, corresponding lines in the optical regime can be detected if they are strong. Detecting elevated N abundances in BSS stars will give a further hint for the binary interaction history of these stars, and provide important insights for binary merger models. Future work also includes a more detailed investigation of the rotational velocities of different MS components, mainly the red and blue MS, which are predicted to be composed of rapidly- and slowly-rotating stars, respectively. Additional constraints can also be provided by the investigation of additional young and intermediate-age clusters in the SMC, LMC and the Milky Way.

\begin{acknowledgements}
 
The authors acknowledge support from the FWO\,Odysseus program under project G0F8H6N and from the European Space Agency (ESA) and the Belgian Federal Science Policy Office (BELSPO) through the PRODEX Programme. The research leading to these results has received funding from the European Research Council (ERC) under the European Union's Horizon 2020 research and innovation programme (grant agreement numbers 772225: MULTIPLES and 945806: TEL-
STARS). 
\end{acknowledgements}

\bibliographystyle{aa}
\bibliography{papers}

\begin{thebibliography}{86}
\expandafter\ifx\csname natexlab\endcsname\relax\def\natexlab#1{#1}\fi

\bibitem[{{Abdul-Masih}(2023)}]{Abdul-Masih2023}
{Abdul-Masih}, M. 2023, \aap, 669, L11

\bibitem[{{Abt} {et~al.}(2002){Abt}, {Levato}, \& {Grosso}}]{Abt2002}
{Abt}, H.~A., {Levato}, H., \& {Grosso}, M. 2002, \apj, 573, 359

\bibitem[{{Aerts} {et~al.}(2019){Aerts}, {Mathis}, \& {Rogers}}]{Aerts2019}
{Aerts}, C., {Mathis}, S., \& {Rogers}, T.~M. 2019, \araa, 57, 35

\bibitem[{{Almeida} {et~al.}(2015){Almeida}, {Sana}, {de Mink}, {Tramper},
  {Soszy{\'n}ski}, {Langer}, {Barb{\'a}}, {Cantiello}, {Damineli}, {de Koter},
  {Garcia}, {Gr{\"a}fener}, {Herrero}, {Howarth}, {Ma{\'\i}z Apell{\'a}niz},
  {Norman}, {Ram{\'\i}rez-Agudelo}, \& {Vink}}]{Almeida2015}
{Almeida}, L.~A., {Sana}, H., {de Mink}, S.~E., {et~al.} 2015, \apj, 812, 102

\bibitem[{{Balona}(1994)}]{Balona1994}
{Balona}, L.~A. 1994, \mnras, 268, 119

\bibitem[{{Bastian} \& {de Mink}(2009)}]{Bastian2009}
{Bastian}, N. \& {de Mink}, S.~E. 2009, \mnras, 398, L11

\bibitem[{{Bodensteiner} {et~al.}(2020){Bodensteiner}, {Sana}, {Mahy},
  {Patrick}, {de Koter}, {de Mink}, {Evans}, {G{\"o}tberg}, {Langer}, {Lennon},
  {Schneider}, \& {Tramper}}]{Bodensteiner2020a}
{Bodensteiner}, J., {Sana}, H., {Mahy}, L., {et~al.} 2020, \aap, 634, A51

\bibitem[{{Bodensteiner} {et~al.}(2021){Bodensteiner}, {Sana}, {Wang},
  {Langer}, {Mahy}, {Banyard}, {de Koter}, {de Mink}, {Evans}, {G{\"o}tberg},
  {Patrick}, {Schneider}, \& {Tramper}}]{Bodensteiner2021}
{Bodensteiner}, J., {Sana}, H., {Wang}, C., {et~al.} 2021, \aap, 652, A70

\bibitem[{{Bouchet} {et~al.}(1985){Bouchet}, {Lequeux}, {Maurice}, {Prevot}, \&
  {Prevot-Burnichon}}]{Bouchet1985}
{Bouchet}, P., {Lequeux}, J., {Maurice}, E., {Prevot}, L., \&
  {Prevot-Burnichon}, M.~L. 1985, \aap, 149, 330

\bibitem[{{Bragan{\c{c}}a} {et~al.}(2012){Bragan{\c{c}}a}, {Daflon}, {Cunha},
  {Bensby}, {Oey}, \& {Walth}}]{Braganca2012}
{Bragan{\c{c}}a}, G.~A., {Daflon}, S., {Cunha}, K., {et~al.} 2012, \aj, 144,
  130

\bibitem[{{Brott} {et~al.}(2011){Brott}, {Evans}, {Hunter}, {de Koter},
  {Langer}, {Dufton}, {Cantiello}, {Trundle}, {Lennon}, {de Mink}, {Yoon}, \&
  {Anders}}]{Brott2011}
{Brott}, I., {Evans}, C.~J., {Hunter}, I., {et~al.} 2011, \aap, 530, A116

\bibitem[{{Carini} {et~al.}(2020){Carini}, {Biazzo}, {Brocato}, {Pulone}, \&
  {Pasquini}}]{Carini2020}
{Carini}, R., {Biazzo}, K., {Brocato}, E., {Pulone}, L., \& {Pasquini}, L.
  2020, \aj, 159, 152

\bibitem[{{Czesla} {et~al.}(2019){Czesla}, {Schr{\"o}ter}, {Schneider},
  {Huber}, {Pfeifer}, {Andreasen}, \& {Zechmeister}}]{pyasl}
{Czesla}, S., {Schr{\"o}ter}, S., {Schneider}, C.~P., {et~al.} 2019, {PyA:
  Python astronomy-related packages}

\bibitem[{{Daflon} {et~al.}(2007){Daflon}, {Cunha}, {de Ara{\'u}jo}, {Wolff},
  \& {Przybilla}}]{Daflon2007}
{Daflon}, S., {Cunha}, K., {de Ara{\'u}jo}, F.~X., {Wolff}, S., \& {Przybilla},
  N. 2007, \aj, 134, 1570

\bibitem[{{de Mink} {et~al.}(2009){de Mink}, {Cantiello}, {Langer}, {Pols},
  {Brott}, \& {Yoon}}]{deMink2009a}
{de Mink}, S.~E., {Cantiello}, M., {Langer}, N., {et~al.} 2009, \aap, 497, 243

\bibitem[{{de Mink} {et~al.}(2013){de Mink}, {Langer}, {Izzard}, {Sana}, \& {de
  Koter}}]{deMink2013}
{de Mink}, S.~E., {Langer}, N., {Izzard}, R.~G., {Sana}, H., \& {de Koter}, A.
  2013, \apj, 764, 166

\bibitem[{{Deb} \& {Singh}(2010)}]{Deb2010}
{Deb}, S. \& {Singh}, H.~P. 2010, \mnras, 402, 691

\bibitem[{{Dufton} {et~al.}(2019){Dufton}, {Evans}, {Hunter}, {Lennon}, \&
  {Schneider}}]{Dufton2019}
{Dufton}, P.~L., {Evans}, C.~J., {Hunter}, I., {Lennon}, D.~J., \& {Schneider},
  F.~R.~N. 2019, \aap, 626, A50

\bibitem[{{Dufton} {et~al.}(2013){Dufton}, {Langer}, {Dunstall}, {Evans},
  {Brott}, {de Mink}, {Howarth}, {Kennedy}, {McEvoy}, {Potter},
  {Ram{\'\i}rez-Agudelo}, {Sana}, {Sim{\'o}n-D{\'\i}az}, {Taylor}, \&
  {Vink}}]{Dufton2013}
{Dufton}, P.~L., {Langer}, N., {Dunstall}, P.~R., {et~al.} 2013, \aap, 550,
  A109

\bibitem[{{Dufton} {et~al.}(2022){Dufton}, {Lennon}, {Villase{\~n}or},
  {Howarth}, {Evans}, {de Mink}, {Sana}, \& {Taylor}}]{Dufton2022}
{Dufton}, P.~L., {Lennon}, D.~J., {Villase{\~n}or}, J.~I., {et~al.} 2022,
  \mnras, 512, 3331

\bibitem[{{Ekstr{\"o}m} {et~al.}(2012){Ekstr{\"o}m}, {Georgy}, {Eggenberger},
  {Meynet}, {Mowlavi}, {Wyttenbach}, {Granada}, {Decressin}, {Hirschi},
  {Frischknecht}, {Charbonnel}, \& {Maeder}}]{Ekstrom2012}
{Ekstr{\"o}m}, S., {Georgy}, C., {Eggenberger}, P., {et~al.} 2012, \aap, 537,
  A146

\bibitem[{{Ekstr{\"o}m} {et~al.}(2008){Ekstr{\"o}m}, {Meynet}, {Maeder}, \&
  {Barblan}}]{Ekstrom2008}
{Ekstr{\"o}m}, S., {Meynet}, G., {Maeder}, A., \& {Barblan}, F. 2008, \aap,
  478, 467

\bibitem[{{Eldridge} {et~al.}(2020){Eldridge}, {Beasor}, \&
  {Britavskiy}}]{Eldridge2020}
{Eldridge}, J.~J., {Beasor}, E.~R., \& {Britavskiy}, N. 2020, \mnras, 495, L102

\bibitem[{{Eldridge} \& {Stanway}(2009)}]{Eldridge2009}
{Eldridge}, J.~J. \& {Stanway}, E.~R. 2009, \mnras, 400, 1019

\bibitem[{{Evans} {et~al.}(2006){Evans}, {Lennon}, {Smartt}, \&
  {Trundle}}]{Evans2006}
{Evans}, C.~J., {Lennon}, D.~J., {Smartt}, S.~J., \& {Trundle}, C. 2006, \aap,
  456, 623

\bibitem[{{Evans} {et~al.}(2011){Evans}, {Taylor}, {H{\'e}nault-Brunet},
  {Sana}, {de Koter}, {Sim{\'o}n-D{\'\i}az}, {Carraro}, {Bagnoli}, {Bastian},
  {Bestenlehner}, {Bonanos}, {Bressert}, {Brott}, {Campbell}, {Cantiello},
  {Clark}, {Costa}, {Crowther}, {de Mink}, {Doran}, {Dufton}, {Dunstall},
  {Friedrich}, {Garcia}, {Gieles}, {Gr{\"a}fener}, {Herrero}, {Howarth},
  {Izzard}, {Langer}, {Lennon}, {Ma{\'\i}z Apell{\'a}niz}, {Markova},
  {Najarro}, {Puls}, {Ramirez}, {Sab{\'\i}n-Sanjuli{\'a}n}, {Smartt}, {Stroud},
  {van Loon}, {Vink}, \& {Walborn}}]{Evans2011}
{Evans}, C.~J., {Taylor}, W.~D., {H{\'e}nault-Brunet}, V., {et~al.} 2011, \aap,
  530, A108

\bibitem[{{Fitzpatrick} {et~al.}(2019){Fitzpatrick}, {Massa}, {Gordon},
  {Bohlin}, \& {Clayton}}]{Fitzpatrick2019}
{Fitzpatrick}, E.~L., {Massa}, D., {Gordon}, K.~D., {Bohlin}, R., \& {Clayton},
  G.~C. 2019, \apj, 886, 108

\bibitem[{{Garland} {et~al.}(2017){Garland}, {Dufton}, {Evans}, {Crowther},
  {Howarth}, {de Koter}, {de Mink}, {Grin}, {Langer}, {Lennon}, {McEvoy},
  {Sana}, {Schneider}, {S{\'\i}mon D{\'\i}az}, {Taylor}, {Thompson}, \&
  {Vink}}]{Garland2017}
{Garland}, R., {Dufton}, P.~L., {Evans}, C.~J., {et~al.} 2017, \aap, 603, A91

\bibitem[{{Garmany} {et~al.}(2015){Garmany}, {Glaspey}, {Bragan{\c{c}}a},
  {Daflon}, {Borges Fernandes}, {Oey}, {Bensby}, \& {Cunha}}]{Garmany2015}
{Garmany}, C.~D., {Glaspey}, J.~W., {Bragan{\c{c}}a}, G.~A., {et~al.} 2015,
  \aj, 150, 41

\bibitem[{{Georgy} {et~al.}(2013){Georgy}, {Ekstr{\"o}m}, {Granada}, {Meynet},
  {Mowlavi}, {Eggenberger}, \& {Maeder}}]{Georgy2013}
{Georgy}, C., {Ekstr{\"o}m}, S., {Granada}, A., {et~al.} 2013, \aap, 553, A24

\bibitem[{{Gordon} {et~al.}(2003){Gordon}, {Clayton}, {Misselt}, {Landolt}, \&
  {Wolff}}]{Gordon2003}
{Gordon}, K.~D., {Clayton}, G.~C., {Misselt}, K.~A., {Landolt}, A.~U., \&
  {Wolff}, M.~J. 2003, \apj, 594, 279

\bibitem[{{Graczyk} {et~al.}(2014){Graczyk}, {Pietrzy{\'n}ski}, {Thompson},
  {Gieren}, {Pilecki}, {Konorski}, {Udalski}, {Soszy{\'n}ski}, {Villanova},
  {G{\'o}rski}, {Suchomska}, {Karczmarek}, {Kudritzki}, {Bresolin}, \&
  {Gallenne}}]{Graczyk2014}
{Graczyk}, D., {Pietrzy{\'n}ski}, G., {Thompson}, I.~B., {et~al.} 2014, \apj,
  780, 59

\bibitem[{{Hastings} {et~al.}(2021){Hastings}, {Langer}, {Wang},
  {Schootemeijer}, \& {Milone}}]{Hastings2021}
{Hastings}, B., {Langer}, N., {Wang}, C., {Schootemeijer}, A., \& {Milone},
  A.~P. 2021, \aap, 653, A144

\bibitem[{{Hastings} {et~al.}(2020){Hastings}, {Wang}, \&
  {Langer}}]{Hastings2020}
{Hastings}, B., {Wang}, C., \& {Langer}, N. 2020, \aap, 633, A165

\bibitem[{{Holgado} {et~al.}(2022){Holgado}, {Sim{\'o}n-D{\'\i}az}, {Herrero},
  \& {Barb{\'a}}}]{Holgado2022}
{Holgado}, G., {Sim{\'o}n-D{\'\i}az}, S., {Herrero}, A., \& {Barb{\'a}}, R.~H.
  2022, \aap, 665, A150

\bibitem[{{Howarth} \& {Smith}(2001)}]{Howarth2001}
{Howarth}, I.~D. \& {Smith}, K.~C. 2001, \mnras, 327, 353

\bibitem[{{Huang} \& {Gies}(2006)}]{Huang2006}
{Huang}, W. \& {Gies}, D.~R. 2006, \apj, 648, 580

\bibitem[{{Huang} \& {Gies}(2008)}]{Huang2008}
{Huang}, W. \& {Gies}, D.~R. 2008, \apj, 683, 1045

\bibitem[{{Huang} {et~al.}(2010){Huang}, {Gies}, \& {McSwain}}]{Huang2010}
{Huang}, W., {Gies}, D.~R., \& {McSwain}, M.~V. 2010, \apj, 722, 605

\bibitem[{{Hubeny} \& {Lanz}(1995)}]{Hubeny1995}
{Hubeny}, I. \& {Lanz}, T. 1995, \apj, 439, 875

\bibitem[{{Hunter} {et~al.}(2007){Hunter}, {Dufton}, {Smartt}, {Ryans},
  {Evans}, {Lennon}, {Trundle}, {Hubeny}, \& {Lanz}}]{Hunter2007}
{Hunter}, I., {Dufton}, P.~L., {Smartt}, S.~J., {et~al.} 2007, \aap, 466, 277

\bibitem[{{Hunter} {et~al.}(2008){Hunter}, {Lennon}, {Dufton}, {Trundle},
  {Sim{\'o}n-D{\'\i}az}, {Smartt}, {Ryans}, \& {Evans}}]{Hunter2008a}
{Hunter}, I., {Lennon}, D.~J., {Dufton}, P.~L., {et~al.} 2008, \aap, 479, 541

\bibitem[{{Kamann} {et~al.}(2018){Kamann}, {Bastian}, {Husser}, {Martocchia},
  {Usher}, {den Brok}, {Dreizler}, {Kelz}, {Krajnovi{\'c}}, {Richard},
  {Steinmetz}, \& {Weilbacher}}]{Kamann2018b}
{Kamann}, S., {Bastian}, N., {Husser}, T.~O., {et~al.} 2018, \mnras, 480, 1689

\bibitem[{{Kamann} {et~al.}(2021){Kamann}, {Bastian}, {Usher}, {Cabrera-Ziri},
  \& {Saracino}}]{Kamann2021}
{Kamann}, S., {Bastian}, N., {Usher}, C., {Cabrera-Ziri}, I., \& {Saracino}, S.
  2021, \mnras, 508, 2302

\bibitem[{{Kamann} {et~al.}(2023){Kamann}, {Saracino}, {Bastian}, {Gossage},
  {Usher}, {Baade}, {Cabrera-Ziri}, {de Mink}, {Ekstrom}, {Georgy}, {Hilker},
  {Larsen}, {Mackey}, {Niederhofer}, {Platais}, \& {Yong}}]{Kamann2023}
{Kamann}, S., {Saracino}, S., {Bastian}, N., {et~al.} 2023, \mnras, 518, 1505

\bibitem[{{Keller} {et~al.}(1999){Keller}, {Wood}, \& {Bessell}}]{Keller1999}
{Keller}, S.~C., {Wood}, P.~R., \& {Bessell}, M.~S. 1999, \aaps, 134, 489

\bibitem[{{Kilian}(1994)}]{Kilian1994}
{Kilian}, J. 1994, \aap, 282, 867

\bibitem[{{Kippenhahn} \& {Weigert}(1967)}]{Kippenhahn1967}
{Kippenhahn}, R. \& {Weigert}, A. 1967, \zap, 65, 251

\bibitem[{{Klencki} {et~al.}(2022){Klencki}, {Istrate}, {Nelemans}, \&
  {Pols}}]{Klencki2022}
{Klencki}, J., {Istrate}, A., {Nelemans}, G., \& {Pols}, O. 2022, \aap, 662,
  A56

\bibitem[{{Kudritzki}(1979)}]{Kudritzki1979}
{Kudritzki}, R.~P. 1979, in Liege International Astrophysical Colloquia,
  Vol.~22, Liege International Astrophysical Colloquia, ed. A.~{Boury},
  N.~{Grevesse}, \& L.~{Remy-Battiau}, 295--318

\bibitem[{{Lanz} \& {Hubeny}(2007)}]{Lanz2007}
{Lanz}, T. \& {Hubeny}, I. 2007, \apjs, 169, 83

\bibitem[{{Lucy}(1974)}]{Lucy1974}
{Lucy}, L.~B. 1974, \aj, 79, 745

\bibitem[{{Maeder} \& {Meynet}(2000)}]{Maeder2000a}
{Maeder}, A. \& {Meynet}, G. 2000, \araa, 38, 143

\bibitem[{{Mahy} {et~al.}(2020){Mahy}, {Sana}, {Abdul-Masih}, {Almeida},
  {Langer}, {Shenar}, {de Koter}, {de Mink}, {de Wit}, {Grin}, {Evans},
  {Moffat}, {Schneider}, {Barb{\'a}}, {Clark}, {Crowther}, {Gr{\"a}fener},
  {Lennon}, {Tramper}, \& {Vink}}]{Mahy2020a}
{Mahy}, L., {Sana}, H., {Abdul-Masih}, M., {et~al.} 2020, \aap, 634, A118

\bibitem[{{Marino} {et~al.}(2018){Marino}, {Przybilla}, {Milone}, {Da Costa},
  {D'Antona}, {Dotter}, \& {Dupree}}]{Marino2018}
{Marino}, A.~F., {Przybilla}, N., {Milone}, A.~P., {et~al.} 2018, \aj, 156, 116

\bibitem[{{Martayan} {et~al.}(2007{\natexlab{a}}){Martayan}, {Floquet},
  {Hubert}, {Guti{\'e}rrez-Soto}, {Fabregat}, {Neiner}, \&
  {Mekkas}}]{Martayan2007b}
{Martayan}, C., {Floquet}, M., {Hubert}, A.~M., {et~al.} 2007{\natexlab{a}},
  \aap, 472, 577

\bibitem[{{Martayan} {et~al.}(2007{\natexlab{b}}){Martayan}, {Fr{\'e}mat},
  {Hubert}, {Floquet}, {Zorec}, \& {Neiner}}]{Martayan2007a}
{Martayan}, C., {Fr{\'e}mat}, Y., {Hubert}, A.~M., {et~al.} 2007{\natexlab{b}},
  \aap, 462, 683

\bibitem[{{Milone} {et~al.}(2018){Milone}, {Marino}, {Di Criscienzo},
  {D'Antona}, {Bedin}, {Da Costa}, {Piotto}, {Tailo}, {Dotter}, {Angeloni},
  {Anderson}, {Jerjen}, {Li}, {Dupree}, {Granata}, {Lagioia}, {Mackey},
  {Nardiello}, \& {Vesperini}}]{Milone2018}
{Milone}, A.~P., {Marino}, A.~F., {Di Criscienzo}, M., {et~al.} 2018, \mnras,
  477, 2640

\bibitem[{{Mokiem} {et~al.}(2007){Mokiem}, {de Koter}, {Vink}, {Puls}, {Evans},
  {Smartt}, {Crowther}, {Herrero}, {Langer}, {Lennon}, {Najarro}, \&
  {Villamariz}}]{Mokiem2007}
{Mokiem}, M.~R., {de Koter}, A., {Vink}, J.~S., {et~al.} 2007, \aap, 473, 603

\bibitem[{{Nieva} \& {Przybilla}(2007)}]{Nieva2007}
{Nieva}, M.~F. \& {Przybilla}, N. 2007, \aap, 467, 295

\bibitem[{{Patrick} {et~al.}(2020){Patrick}, {Lennon}, {Evans}, {Sana},
  {Bodensteiner}, {Britavskiy}, {Dorda}, {Herrero}, {Negueruela}, \& {de
  Koter}}]{Patrick2020}
{Patrick}, L.~R., {Lennon}, D.~J., {Evans}, C.~J., {et~al.} 2020, \aap, 635,
  A29

\bibitem[{{Pols} {et~al.}(1991){Pols}, {Cote}, {Waters}, \& {Heise}}]{Pols1991}
{Pols}, O.~R., {Cote}, J., {Waters}, L.~B.~F.~M., \& {Heise}, J. 1991, \aap,
  241, 419

\bibitem[{{Ramachandran} {et~al.}(2018){Ramachandran}, {Hamann}, {Hainich},
  {Oskinova}, {Shenar}, {Sander}, {Todt}, \& {Gallagher}}]{Ramachandran2018a}
{Ramachandran}, V., {Hamann}, W.~R., {Hainich}, R., {et~al.} 2018, \aap, 615,
  A40

\bibitem[{{Ramachandran} {et~al.}(2019){Ramachandran}, {Hamann}, {Oskinova},
  {Gallagher}, {Hainich}, {Shenar}, {Sander}, {Todt}, \&
  {Fulmer}}]{Ramachandran2019}
{Ramachandran}, V., {Hamann}, W.~R., {Oskinova}, L.~M., {et~al.} 2019, \aap,
  625, A104

\bibitem[{{Ram{\'\i}rez-Agudelo} {et~al.}(2015){Ram{\'\i}rez-Agudelo}, {Sana},
  {de Mink}, {H{\'e}nault-Brunet}, {de Koter}, {Langer}, {Tramper},
  {Gr{\"a}fener}, {Evans}, {Vink}, {Dufton}, \& {Taylor}}]{Ramirez-Agudelo2015}
{Ram{\'\i}rez-Agudelo}, O.~H., {Sana}, H., {de Mink}, S.~E., {et~al.} 2015,
  \aap, 580, A92

\bibitem[{{Ram{\'\i}rez-Agudelo} {et~al.}(2013){Ram{\'\i}rez-Agudelo},
  {Sim{\'o}n-D{\'\i}az}, {Sana}, {de Koter}, {Sab{\'\i}n-Sanjul{\'\i}an}, {de
  Mink}, {Dufton}, {Gr{\"a}fener}, {Evans}, {Herrero}, {Langer}, {Lennon},
  {Ma{\'\i}z Apell{\'a}niz}, {Markova}, {Najarro}, {Puls}, {Taylor}, \&
  {Vink}}]{Ramirez-Agudelo2013}
{Ram{\'\i}rez-Agudelo}, O.~H., {Sim{\'o}n-D{\'\i}az}, S., {Sana}, H., {et~al.}
  2013, \aap, 560, A29

\bibitem[{{Rivinius} {et~al.}(2003){Rivinius}, {Baade}, \&
  {{\v{S}}tefl}}]{Rivinius2003}
{Rivinius}, T., {Baade}, D., \& {{\v{S}}tefl}, S. 2003, \aap, 411, 229

\bibitem[{{Rivinius} {et~al.}(2013){Rivinius}, {Carciofi}, \&
  {Martayan}}]{Rivinius2013}
{Rivinius}, T., {Carciofi}, A.~C., \& {Martayan}, C. 2013, \aapr, 21, 69

\bibitem[{{Rogers} {et~al.}(2013){Rogers}, {Lin}, {McElwaine}, \&
  {Lau}}]{Rogers2013}
{Rogers}, T.~M., {Lin}, D.~N.~C., {McElwaine}, J.~N., \& {Lau}, H.~H.~B. 2013,
  \apj, 772, 21

\bibitem[{{Schneider} {et~al.}(2020){Schneider}, {Ohlmann}, {Podsiadlowski},
  {R{\"o}pke}, {Balbus}, \& {Pakmor}}]{Schneider2020}
{Schneider}, F.~R.~N., {Ohlmann}, S.~T., {Podsiadlowski}, P., {et~al.} 2020,
  \mnras, 495, 2796

\bibitem[{{Schneider} {et~al.}(2018){Schneider}, {Ram{\'\i}rez-Agudelo},
  {Tramper}, {Bestenlehner}, {Castro}, {Sana}, {Evans},
  {Sab{\'\i}n-Sanjuli{\'a}n}, {Sim{\'o}n-D{\'\i}az}, {Langer}, {Fossati},
  {Gr{\"a}fener}, {Crowther}, {de Mink}, {de Koter}, {Gieles}, {Herrero},
  {Izzard}, {Kalari}, {Klessen}, {Lennon}, {Mahy}, {Ma{\'\i}z Apell{\'a}niz},
  {Markova}, {van Loon}, {Vink}, \& {Walborn}}]{Schneider2018}
{Schneider}, F.~R.~N., {Ram{\'\i}rez-Agudelo}, O.~H., {Tramper}, F., {et~al.}
  2018, \aap, 618, A73

\bibitem[{{Sim{\'o}n-D{\'\i}az} {et~al.}(2017){Sim{\'o}n-D{\'\i}az}, {Godart},
  {Castro}, {Herrero}, {Aerts}, {Puls}, {Telting}, \&
  {Grassitelli}}]{Simon-Diaz2017}
{Sim{\'o}n-D{\'\i}az}, S., {Godart}, M., {Castro}, N., {et~al.} 2017, \aap,
  597, A22

\bibitem[{{Sim{\'o}n-D{\'\i}az} \& {Herrero}(2014)}]{Simon-Diaz2014}
{Sim{\'o}n-D{\'\i}az}, S. \& {Herrero}, A. 2014, \aap, 562, A135

\bibitem[{{Slettebak}(1949)}]{Slettebak1949}
{Slettebak}, A. 1949, \apj, 110, 498

\bibitem[{{Slettebak} {et~al.}(1975){Slettebak}, {Collins}, {Boyce}, {White},
  \& {Parkinson}}]{Slettebak1975}
{Slettebak}, A., {Collins}, G.~W., I., {Boyce}, P.~B., {White}, N.~M., \&
  {Parkinson}, T.~D. 1975, \apjs, 29, 137

\bibitem[{{Steele} {et~al.}(1999){Steele}, {Negueruela}, \&
  {Clark}}]{Steele1999}
{Steele}, I.~A., {Negueruela}, I., \& {Clark}, J.~S. 1999, \aaps, 137, 147

\bibitem[{{Strom} {et~al.}(2005){Strom}, {Wolff}, \& {Dror}}]{Strom2005}
{Strom}, S.~E., {Wolff}, S.~C., \& {Dror}, D. H.~A. 2005, \aj, 129, 809

\bibitem[{{Sun} {et~al.}(2021){Sun}, {de Grijs}, {Deng}, \& {Albrow}}]{Sun2021}
{Sun}, W., {de Grijs}, R., {Deng}, L., \& {Albrow}, M.~D. 2021, \mnras, 502,
  4350

\bibitem[{Townsend {et~al.}(2004)Townsend, Owocki, \& Howarth}]{Townsend2004}
Townsend, R. H.~D., Owocki, S.~P., \& Howarth, I.~D. 2004, MNRAS, 350, 189

\bibitem[{{von Zeipel}(1924)}]{VonZeipel1924}
{von Zeipel}, H. 1924, \mnras, 84, 665

\bibitem[{{Wang} {et~al.}(2022{\natexlab{a}}){Wang}, {Hastings},
  {Schootemeijer}, {Langer}, {de Mink}, {Bodensteiner}, {Milone}, {Justham}, \&
  {Marchant}}]{WangC2022}
{Wang}, C., {Hastings}, B., {Schootemeijer}, A., {et~al.} 2022{\natexlab{a}},
  arXiv e-prints, arXiv:2211.15794

\bibitem[{{Wang} {et~al.}(2020){Wang}, {Langer}, {Schootemeijer}, {Castro},
  {Adscheid}, {Marchant}, \& {Hastings}}]{WangC2020}
{Wang}, C., {Langer}, N., {Schootemeijer}, A., {et~al.} 2020, \apjl, 888, L12

\bibitem[{{Wang} {et~al.}(2022{\natexlab{b}}){Wang}, {Langer}, {Schootemeijer},
  {Milone}, {Hastings}, {Xu}, {Bodensteiner}, {Sana}, {Castro}, {Lennon},
  {Marchant}, {Koter}, \& {Mink}}]{WangC2021}
{Wang}, C., {Langer}, N., {Schootemeijer}, A., {et~al.} 2022{\natexlab{b}},
  Nature Astronomy, 6, 480

\bibitem[{{Wolff} {et~al.}(2007){Wolff}, {Strom}, {Dror}, \&
  {Venn}}]{Wolff2007}
{Wolff}, S.~C., {Strom}, S.~E., {Dror}, D., \& {Venn}, K. 2007, \aj, 133, 1092

\bibitem[{{Woosley} \& {Heger}(2006)}]{Woosley2006}
{Woosley}, S.~E. \& {Heger}, A. 2006, \apj, 637, 914

\bibitem[{{Zorec} {et~al.}(2016){Zorec}, {Fr{\'e}mat}, {Domiciano de Souza},
  {Royer}, {Cidale}, {Hubert}, {Semaan}, {Martayan}, {Cochetti}, {Arias},
  {Aidelman}, \& {Stee}}]{Zorec2016}
{Zorec}, J., {Fr{\'e}mat}, Y., {Domiciano de Souza}, A., {et~al.} 2016, \aap,
  595, A132

\end{thebibliography}

\begin{appendix}
\section{Stellar parameters derived in this work}\label{app:tab_allparams}
Table~\ref{tab:all_params} gives an overview over the derived stellar parameters for each star with respective errors, namely the effective temperature, surface gravity, projected rotational velocity, radius and luminosity. We also repeat here the spectral types and binary classification from \citetalias{Bodensteiner2020a} and \citetalias{Bodensteiner2021}.

\begin{table*}
\renewcommand{\arraystretch}{1.3}
    \caption{\label{tab:all_params} Compilation of all parameters derived for the sample stars (extract).}
\begin{tabular}{llllllllllll} \\ \hline \hline
Ident & RA & DEC & F336W & F814W & SpT & \Teff & \logg & \vsini &  $R$ & $L$ & Flag \\
BSM & [J2000] & [J2000] & [mag] & [mag] & & [K] & [dex] & [\kms] & [\Rsun] & [\Lsun] & \\ \hline
010 & 14.078913 & -72.472822 & 16.8 & 17.8 & B3e & $16000^{+600}_{-800}$ & $3.2^{+0.5}_{-...}$ & $200^{+50}_{-50}$ & $4.3^{+0.1}_{-0.2}$ & $3.0^{+0.1}_{-0.1}$ &  \\
020 & 14.053745 & -72.472215 & 16.5 & 18.0 & B2 & $23000^{+900}_{-700}$ & $4.2^{+0.2}_{-0.2}$ & $80^{+60}_{-...}$ & $3.0^{+0.0}_{-0.1}$ & $3.4^{+0.1}_{-0.1}$ &  \\
021 & 14.098794 & -72.471839 & 15.7 & 17.2 & B1 & $22000^{+2600}_{-3300}$ & $3.8^{+0.3}_{-0.4}$ & $0^{+100}_{-...}$ & $4.6^{+1.0}_{-0.7}$ & $3.6^{+0.4}_{-0.4}$ & SB1 \\
023 & 14.071630 & -72.471714 & 16.4 & 17.7 & B4 & $18000^{+700}_{-400}$ & $4.0^{+0.2}_{-0.1}$ & $140^{+40}_{-50}$ & $4.3^{+0.2}_{-0.3}$ & $3.2^{+0.1}_{-0.1}$ &  \\
025 & 14.087501 & -72.471613 & 17.2 & 18.4 & B6 & $18000^{+700}_{-1900}$ & $3.8^{+0.1}_{-0.3}$ & $420^{+70}_{-80}$ & $3.0^{+0.3}_{-0.1}$ & $2.9^{+0.2}_{-0.2}$ &  \\
030 & 14.093453 & -72.471074 & 15.4 & 16.5 & B5e & $17000^{+2100}_{-1300}$ & $3.0^{+0.5}_{-...}$ & $320^{+90}_{-80}$ & $7.4^{+0.8}_{-1.2}$ & $3.6^{+0.3}_{-0.3}$ &  \\
037 & 14.065186 & -72.470717 & 16.6 & 18.0 & B6 & $20000^{+500}_{-1900}$ & $4.0^{+0.1}_{-0.2}$ & $160^{+40}_{-60}$ & $3.3^{+0.4}_{-0.1}$ & $3.2^{+0.1}_{-0.2}$ &  \\
041 & 14.044353 & -72.470482 & 14.7 & 16.0 & B4 & $19000^{+400}_{-900}$ & $3.2^{+0.2}_{-0.1}$ & $140^{+20}_{-30}$ & $8.9^{+0.9}_{-0.9}$ & $4.0^{+0.1}_{-0.2}$ &  \\
042 & 14.077542 & -72.470423 & 16.4 & 17.9 & B3 & $20000^{+2200}_{-900}$ & $4.0^{+0.3}_{-0.1}$ & $0^{+60}_{-...}$ & $3.7^{+0.2}_{-0.5}$ & $3.3^{+0.2}_{-0.2}$ &  \\
043 & 14.095624 & -72.470412 & 15.9 & 17.4 & B2 & $20000^{+1400}_{-700}$ & $3.8^{+0.2}_{-0.2}$ & $120^{+40}_{-30}$ & $4.6^{+0.6}_{-0.8}$ & $3.5^{+0.2}_{-0.2}$ &  \\
...\\
\hline
\end{tabular}
\tablefoot{The first, second, and third columns list the identifier as well as coordinates, while the next two columns give the F336W and F814W magnitudes. The spectral type reported in \citetalias{Bodensteiner2020a} is given in column 'SpT'. The next five columns give the main derived stellar parameters (\Teff, \logg, \vsini, $R$, and $L$) with associated errors ('\textit{...}' indicates that the error bar hits the edge of our model grid). The last column gives a binary flag for stars detected as binary candidates in \citetalias{Bodensteiner2021} ('SB1' implies a star was flagged as binary based on RV variability while 'SB2' indicates a composite SB2 spectrum). The full version of this table is available at the CDS. The first few lines are shown as an example.}
\end{table*}

\section{A second test case: the Be star \#688}\label{app:test688}
\begin{figure} \centering
    \includegraphics[width=0.99\hsize]{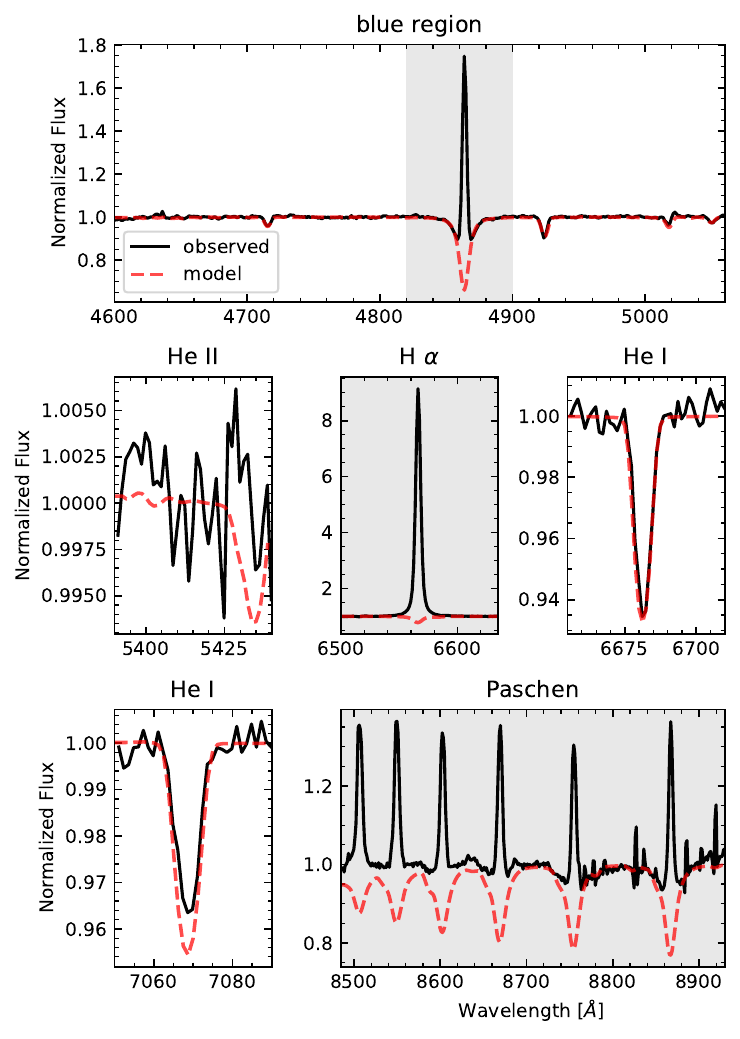}
    \caption{\label{Fig:specfit_id688} Spectroscopic fit for star \#~688, similar to Fig.\,\ref{Fig:specfit_id654} for \#~654. Each panel, one for each of the six diagnostic spectral regions, shows the combined spectrum (black) and the best-fitting \textsc{tlusty} model (red). Grayed-out regions are only shown for comparison but were not included in the fit.}
\end{figure}

\begin{figure} \centering
    \includegraphics[width=0.99\hsize]{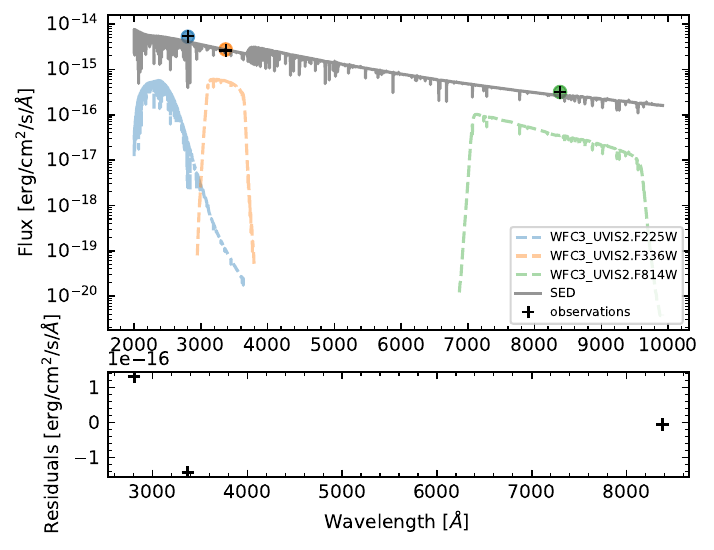}
    \caption{\label{Fig:photfit_id688} Photometric fit for star \#688, similar to Fig.\,\ref{Fig:photfit_id654} for \#654. The top panel shows a comparison between the observed HST fluxes (black crosses) and the flux (colored circles) computed from the best-fit model (gray line) by convolution with the HST filters (colored lines). The bottom panel gives residuals.}
\end{figure}

\begin{figure} \centering
    \includegraphics[width=0.99\hsize]{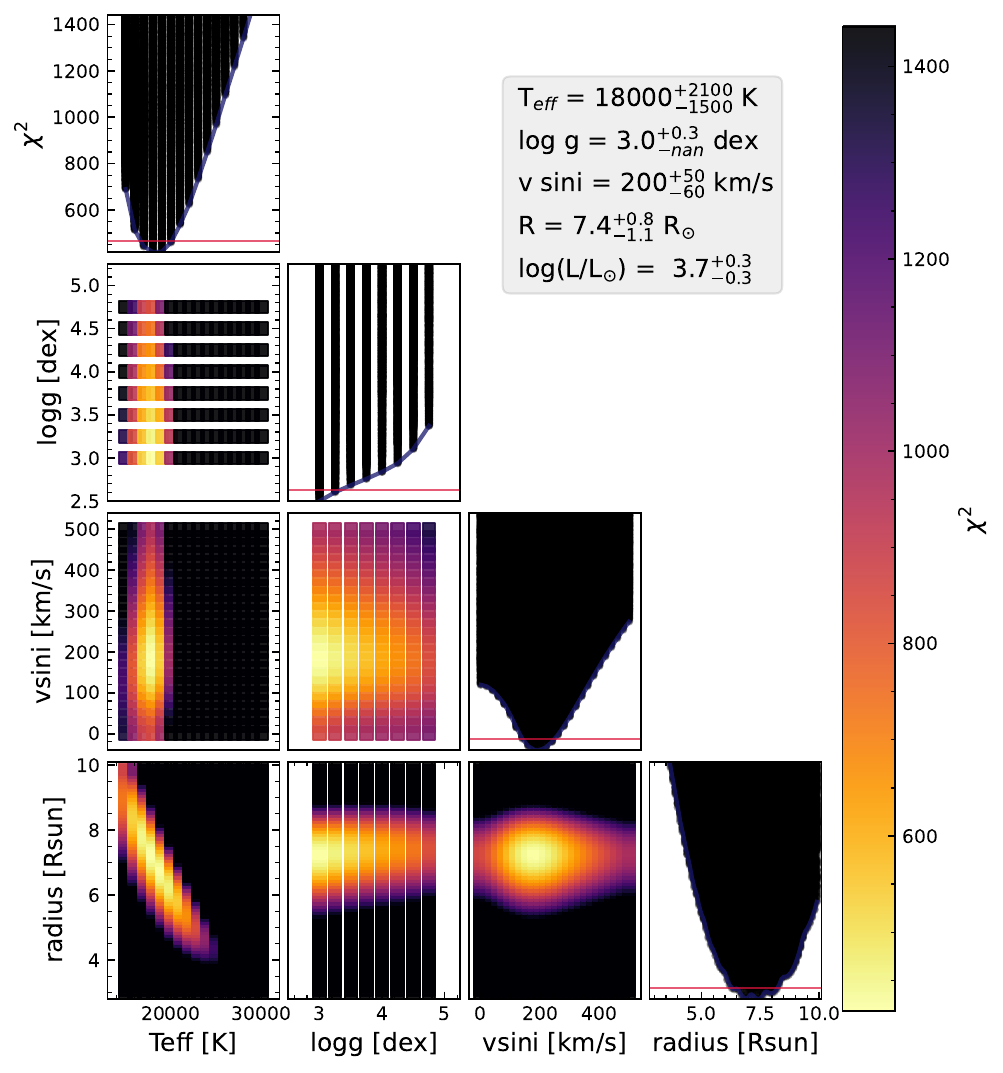}
    \caption{\label{fig:chi2_combined_688} Combined fit for star \#688, similar to Fig.\,\ref{fig:chi2_combined_654} for \#654. The diagonal gives the $\chi^2$-distribution as a function of each parameter and the panels below show 2-dimensional $\chi^2$-maps. The red line marks the 95\% confidence level.}
\end{figure}

Given that our test case \#654 was a slowly rotating B-type star, we here show a second star, a rapidly rotating Be star, namely \#688 to demonstrate how the method works for rapid rotators. Figures\,\ref{Fig:specfit_id688}, \ref{Fig:photfit_id688}, and \ref{fig:chi2_combined_688} are analogous to Figures \ref{Fig:specfit_id654}, \ref{Fig:photfit_id654}, and \ref{fig:chi2_combined_654} for \#654.

Given the strong contamination of the spectrum by emission, the affected lines (in this case mainly H$\beta$, are excluded from the fit). However, the fit of the \ion{He}{i} line at 7065.19\,$\AA$ might imply that there is some infilling as well. The \halpha\, and Paschen lines show strong contamination, but those lines are excluded from the fit for all stars (see Sect.\,\ref{subsec:specfit}). Whether the photometry is impacted by the Be disk is hard to assess from three photometric points only. Given that the HST observations were not taken simultaneously, we cannot extrapolate from the MUSE spectra either.

The combined fit suggests that the Be star \#~688 has the following parameters with uncertainties: \Teff\,=\,18000$^{+2100}_{-1500}$\,K, \logg\,=\,3.0$^{+0.3}_{-...}$\,dex, \vsini\,=\,200$^{+50}_{-60}$\,\kms, $R$\,=\,7.4$^{+0.8}_{-1.1}$\,\Rsun, and $\log$(L/\Lsun)\,=\,$3.7^{+0.3}_{-0.3}$\,\Lsun. As the best-fit value for the surface gravity hits the edge of the grid, no lower error can be given. On average, the error bars are higher than the ones determined for most B-type stars. This is due to the fact that fewer spectral lines are used in the fit. The correlation plot in Fig.\,\ref{fig:chi2_combined_688} further demonstrates that the \logg\, is not well constrained when no Balmer lines (i.e., in particular H$\beta$ in our case) can be included in the fit.


\section{Stellar parameters from \citet{Carini2020}}
Based on one epoch of MUSE-WFM-NOAO commissioning data obtained with the nominal wavelength range in August 2014, \citet{Carini2020} determine Johnson V and Cousins I magnitudes from the MUSE data and extract spectra via PSF for 10 bright B-type stars in the core of NGC\,330. Using photometric estimates as input, they compare the MUSE spectroscopy in the wavelength range from $\lambda\lambda$ 4800-5000$\AA$ with synthetic models to estimate stellar parameters and measured He abundances. In the following, we compare our derived parameters with the spectroscopic parameters provided by \citet{Carini2020}, noting that their average S/N is slightly lower than the one used in this study. Furthermore, given that the observations were taken without AO support and with an average seeing of 1.3", crowding is expected to be a more severe problem here.

Given a similar pointing, all of the ten stars are included in our study. Two of the ten stars, namely \#313 (A10) and \#519 (A48) were classified as SB1s in \citetalias{Bodensteiner2021}. Because we do not detect any signature of a companion in their spectrum we fit the spectrum like the one of a single star. The comparison of the derived parameters (see Tab.\,\ref{Tab:comp_carini}) shows that they are in reasonable agreement. 

\begingroup
\renewcommand{\arraystretch}{1.3}
\begin{table*}[]\centering
  \caption{Comparison to parameters determined in \citet{Carini2020} of the ten stars in common between the studies.}
    \begin{tabular}{l l l l l  l l l l l} \hline \hline
       \multicolumn{5}{c}{this work} & \multicolumn{4}{c}{\citet{Carini2020}} \\ \hline
        ID & \Teff & \logg & \vsini & Flag & & ID & \Teff & \logg & \vsini \\
        & [K] & [dex] & [\kms\,]  & &  & & [K] & [dex] & [\kms\,] \\ \hline
        670  & $18000^{+400}_{-900}$ & $3.0^{+0.1}_{-...}$ & $20^{+50}_{-...}$ & & & A4 & 17000$\pm$1000 & 2.9$\pm$0.1 & 100 \\
        643 & $21000^{+300}_{-600}$ & $4.0^{+0.1}_{-0.1}$ & $240^{+30}_{-30}$ & & & A5 & 21400$\pm$3000 & 3.5$\pm$0.2 & 250 \\
        313 & $19000^{+800}_{-400}$ & $3.2^{+0.2}_{-0.2}$ & $220^{+30}_{-30}$ & SB1 & & A10 & 20500$\pm$1000 & 3.7$\pm$0.1 & 200 \\
        289 & $20000^{+3300}_{-900}$ & $3.2^{+0.4}_{-0.2}$ & $0^{+80}_{-30}$ & & & A17 & 21700$\pm$1000 & 3.6$\pm$0.1 & 100 \\
        420 & $30000^{+...}_{-600}$ & $3.0^{+0.1}_{-...}$ & $180^{+100}_{-120}$ & & & A21 & 21000$\pm$1000 & 3.6$\pm$0.2 & 150 \\
        522 & $25000^{+1000}_{-3200}$ & $3.8^{+0.1}_{-0.4}$ & $0^{+50}_{...}$ & & & A23 & 21600$\pm$2000 & 3.6$\pm$0.2 & 100 \\
        573 & $19000^{+600}_{-400}$ & $3.5^{+0.1}_{-0.1}$ & $140^{+20}_{-30}$ & & & A24 & 21700$\pm$1500 & 3.9$\pm$0.3 & 150 \\
        134 & $21000^{+300}_{-100}$ & $3.5^{+0.1}_{-0.2}$ & $120^{+200}_{-...}$ & & & A34 & 20500$\pm$1500 & 3.8$\pm$0.2 & 150  \\
        228 & $17000^{+200}_{-700}$ & $3.0^{+0.1}_{-...}$ & $0^{+30}_{-...}$ & & & A40 & 17500$\pm$500 & 3.1$\pm$0.1 & 100 \\
        519 & $29000^{+...}_{-1600}$ & $3.8^{+0.2}_{-0.2}$ & $200^{+50}_{-40}$ & SB1 & & A48 & 22000$\pm$1500 & 3.1$\pm$0.3 & 150  \\
        \hline
    \end{tabular}
    \label{Tab:comp_carini}
    \tablefoot{Here, we use the parameters given in table 4 of \citet{Carini2020} which are based on their analysis considering stellar rotation. The two SB1s detected in \citetalias{Bodensteiner2021} are indicated in the column Flag.}
\end{table*}
\endgroup

The largest discrepancy in the estimated temperatures and surface gravities is found for the SB1 \#519 which might imply that there is some contamination of the secondary that manifests itself differently in the two different fitting processes, for example due to the different wavelength region considered. \#420 also shows a large discrepancy in the derived parameters. This might be due to the fact that \#420 is a Be star with strong emission line contamination in the diagnostic spectral lines, and might imply that at least for those stars, the errors are underestimated. Given that a strong \ion{He}{ii} absorption line is present in the spectrum of \#420 the values derived in this work seem more plausible. We further note that the temperature estimated for the third hottest star, \#522, is also significantly lower in \citet{Carini2020}. The temperature estimates of all three stars could also be impacted by the fact that no \ion{He}{ii} was included in the analysis of \citet{Carini2020}.

While \citet{Carini2020} only considered rotational velocities above 100\,\kms\, in steps of 50\,\kms, the derived rotational velocities follow a similar trend. They furthermore do not provide any uncertainties on \vsini, which hampers the comparison.

\end{appendix}
\end{document}